\documentclass[a4paper,11pt]{article}
\pdfoutput=1 % if your are submitting a pdflatex (i.e. if you have
\usepackage{jheppub} 
\usepackage{multirow}
\usepackage{subfig}

\usepackage[T1]{fontenc}
\usepackage{comment}
\newcommand{\as}{\alpha_s}
\newcommand{\LamQCD}{\Lambda_{\rm QCD}}
\newcommand{\bLam}{\bar{\Lambda}}
\def \be{\begin{equation}}
\def \ee{\end{equation}}
\newcommand{\bea}{\begin{eqnarray}}
\newcommand{\eea}{\end{eqnarray}}

\newcommand{\mupi}{\mu_\pi^2}
\newcommand{\mug}{\mu_G^2}
\newcommand{\rd}{\rho_D^3}
\newcommand{\rls}{\rho_{LS}^3}

\title{\boldmath The $q^2$ moments in inclusive semileptonic $B$ decays}
\author[a]{G. Finauri}
\author[a,b,c]{P. Gambino}

\affiliation[a]{Technische Universit{\"a}t M{\"u}nchen, Physik Department T31, \\ James-Franck-Stra{\ss}e 1, 85748 Garching, Germany}
\affiliation[b]{Dip.\ di Fisica, Universit\`a di Torino \& INFN, Sezione di Torino\\
via Giuria 1, 10125 Torino,
Italy }
\affiliation[c]{Max Planck Institute for Physics, \\ F{\"o}hringer Ring 6, 80805 M{\"u}nchen, Germany}

\emailAdd{gael.finauri@tum.de, gambino@to.infn.it}

\abstract{We compute the first moments of the $q^2$ distribution in inclusive semileptonic $B$ decays as functions of the lower cut on $q^2$, confirming a number of results given in the literature and adding the $O(\alpha_s^2\beta_0)$ BLM contributions. We then include the $q^2$-moments recently measured by Belle and Belle II 
in a global fit to the moments. The new data are compatible with the other measurements and slightly decrease the 
uncertainty on the nonperturbative parameters and on $|V_{cb}|$. Our updated value is $|V_{cb}|=(41.97\pm 0.48)\times 10^{-3}$.}
\begin{document} 
\begin{flushright}
{\small
TUM-HEP-1477/23\\
October 31, 2023
%\\arXiv:20mm.nnnnn [hep-ph]
}
\end{flushright}
\maketitle
\flushbottom

\section{Introduction}
\label{sec:intro}
The measurement of the first few moments of the kinematic distributions in inclusive semileptonic $B$ decays has been instrumental in  determining  the CKM matrix element $|V_{cb}|$ from these decays. For over 20 years the moments of the lepton energy  and hadronic invariant mass distributions measured by CLEO, DELPHI, CDF, BaBar and Belle \cite{BaBar:2004bij,CLEO:2004bqt,CDF:2005xlh,DELPHI:2005mot,Belle:2006jtu,Belle:2006kgy,BaBar:2009zpz} have been employed in global fits \cite{Bauer:2004ve,Buchmuller:2005zv,Gambino:2013rza,Alberti:2014yda, Bordone:2021oof,HeavyFlavorAveragingGroup:2022wzx} to extract $|V_{cb}|$. The fits are 
based on the Operator Product Expansion (OPE) which governs these inclusive decays, and  use the moments to get information on the  nonperturbative parameters of the OPE, namely the matrix elements of the local operators. Thanks to a sustained effort in computing higher order effects, which culminated in the $O(\alpha_s^3)$
calculation of the semileptonic width \cite{Fael:2020tow}, the  inclusive determination of   $|V_{cb}|$ has currently a 1.2\% uncertainty \cite{Bordone:2021oof}.

The moments of the lepton invariant mass ($q^2$) distribution have been recently proposed as a basis   to extract $|V_{cb}|$ with a 
reduced set of higher dimensional  parameters \cite{Fael:2018vsp}. Although a couple of $q^2$ moments had been measured by CLEO long ago, they were subject to  a lower cut on the lepton energy and therefore unsuitable to the  method proposed in \cite{Fael:2018vsp}, which relies on Reparametrization Invariance (RPI) \cite{Luke:1992cs,Mannel:2010wj} relations and  requires instead a lower cut on $q^2$. 
This has led to  new and  precise measurements of the first few $q^2$ moments by the  Belle and Belle II collaborations \cite{Belle:2021idw,Belle-II:2022fug}, which were then employed in \cite{Bernlochner:2022ucr}, together with additional information,  to extract  $|V_{cb}|$ with the RPI method of Ref.~\cite{Fael:2018vsp}. 
The purpose of this paper is twofold: first, we want to revisit the OPE calculation of the $q^2$ moments, checking and extending the results that can be found in the literature; second, we want to verify the 
compatibility of these new measurements with previous results for leptonic and hadronic moments, and to study 
their impact on the global OPE fit and the determination of $|V_{cb}|$.

As we plan to make use of all the available inclusive measurements, 
we perform our calculation  of the $q^2$ moments
in the kinetic scheme framework introduced in \cite{Gambino:2004qm} and later adopted and developed in \cite{Gambino:2013rza,Alberti:2014yda,Bordone:2021oof} and do not  employ the RPI approach of \cite{Fael:2018vsp}. 
We compute the first three moments of the $q^2$ spectrum subject to  a lower cut on $q^2$, including all available higher order effects. In particular we compute  power corrections up to $O(1/m_b^3)$ starting from the structure functions given in \cite{Blok:1993va,Manohar:1993qn,Gremm:1996df}. The 
 $O(1/m_b^4)$,  $O(1/m_b^5)$,  and  $O(1/(m_c^2 m_b^3))$ effects depend on a number of additional nonperturbative parameters and are available for the leptonic and hadronic moments \cite{Mannel:2010wj,Gambino:2016jkc}, but would require a dedicated calculation which we postpone to a future publication. We also include the $O(\alpha_s)$ and 
    $O(\alpha_s^2\beta_0)$ perturbative corrections  starting from the triple differential distribution given in \cite{Aquila:2005hq,Trott:2004xc} and \cite{Aquila:2005hq}, respectively. Finally, we include all the  $O(\alpha_s/m_b^2)$ and the $O(\alpha_s \rho_D^3/m_b^3)$ and $O(\alpha_s \rho_{LS}^3/m_b^3)$  contribution using Refs.~\cite{Alberti:2012dn,Alberti:2013kxa} and \cite{Mannel:2021zzr} and RPI, respectively. Additional $O(\alpha_s/m_b^3)$ corrections might  be expected from four-quark operators \cite{Manohar:2010sf}.
    We find perfect agreement with the  results given in 
\cite{Fael:2018vsp,Fael:2020tow,Mannel:2021zzr}, but the $O(\alpha_s^2\beta_0)$ perturbative corrections to the $q^2$ moments are presented here for the first time.
They are expected to provide the bulk of the two-loop corrections.

As they do not depend on $\mu_\pi^2$, the $q^2$ moments can play an important role in a global fit to inclusive data. 
Indeed, fits based only on the leptonic and hadronic moments  find a high correlation between  the matrix elements 
$\mu_\pi^2$ and $\rho_D^3$, 0.73 in the default fit of \cite{Bordone:2021oof}. As they probe a new direction in the 
parameter space, the inclusion of the $q^2$ moments in the fit can  improve the determination of the OPE 
parameters and  decrease the uncertainty of the inclusive determination of $|V_{cb}|$. Moreover, the Belle and Belle II 
measurements
\cite{Belle:2021idw,Belle-II:2022fug} are the first measurements  of inclusive semileptonic $B$ decays since 2009, and it  is  important to test their compatibility 
with older measurements of leptonic and hadronic moments. Indeed, the value of $\rho_D^3(1~{\rm GeV})$ obtained 
in \cite{Bordone:2021oof}, $0.185\pm0.031~\text{GeV}^3$, is in strong tension with the one extracted in \cite{Bernlochner:2022ucr} in a fit including only terms up to $O(1/m_b^3)$, $0.03\pm 0.02~\text{GeV}^3$, suggesting that the $q^2$ moments may be incompatible with previous measurements. Even though the fits of Refs.~\cite{Bordone:2021oof,Bernlochner:2022ucr} employ
slightly different conventions for the OPE parameters, and therefore the
two above values of $\rho_D^3$ should not be compared directly, and even though the  results for $|V_{cb}|$  seem unaffected by this discrepancy, the situation needs to be clarified. 

This paper is organized as follows. In Sec.~\ref{sec:calc} we describe our calculation of the $q^2$ moments and compare our results with the Belle and Belle II measurements. In Sec.~\ref{sec:fit} we discuss the inclusion of these measurements in the global fit to inclusive semileptonic data performed in \cite{Bordone:2021oof}. Sec.~\ref{sec:conc} contains our conclusions.

\section{The calculation}
\label{sec:calc}
The kinematics of inclusive semileptonic $B$ decays is described by three independent variables. 
We work in the $\bar{B}$ meson rest frame, with momentum $p_B = m_B v$, with $v = (1,0,0,0)$. We will denote with $p_X$ the momentum of the hadronic state and with $q$ the momentum of the lepton-neutrino pair. Momentum conservation implies
\begin{equation}
q= p_B - p_X =p_b -p_h\, , \label{eq:q}
\end{equation}
where $p_b$ and $p_h$ are the momenta of the $b$ quark and the partonic inclusive final state, respectively.
We choose as independent kinematical variables  the dilepton invariant mass $q^2$, the charged lepton energy $v \cdot p_\ell = E_\ell$, and the charm off-shellness $u$ defined as
\begin{equation}
u = (p_b-q)^2 - m_c^2\,.
\end{equation}
We will mostly work with the dimensionless quantities 
\begin{equation}
\rho = \frac{m_c^2}{m_b^2}\,, \qquad \hat{u} = \frac{u}{m_b^2}\,, \qquad\hat{E}_\ell = \frac{E_\ell}{m_b}\,, \qquad\hat{q}^2 = \frac{q^2}{m_b^2}\, .
\end{equation}
In the absence of cuts on the lepton energy, the kinematical boundaries on these variables are
\begin{align}
\frac{\hat{q}_0 - \sqrt{\hat{q}_0^2-\hat{q}^2}}{2} \leq\, &\hat{E}_\ell \leq \frac{\hat{q}_0 + \sqrt{\hat{q}_0^2-\hat{q}^2}}{2}\,,\\
0 \leq\, &\hat{u} \leq \hat{u}_+ = (1-\sqrt{\hat{q}^2})^2-\rho\,,\\
0 \leq\, &\hat{q}^2 \leq (1-\sqrt{\rho})^2\,,
\label{eq:q2lims}
\end{align}
where $\hat{q}_0 = (1+\hat{q}^2-\rho-\hat{u})/2$.
We now assume\footnote{The $b$ quark momentum is in general $p_b = m_b v + k$. However one could define the partonic inclusive momentum to be $p_h \equiv m_b v - q$, effectively absorbing $k$ into $p_h$, without changing the rest of the analysis.} $p_b = m_b v$ and introduce
\begin{equation}
    \bLam = m_B - m_b\,, \qquad m_X^2 = p_h^2, \qquad \varepsilon_X = v \cdot p_h\, .
\end{equation}
The hadronic invariant mass 
\begin{equation}
    M_X^2 = (p_B-q)^2 = m_B^2 -2m_B q_0 +q^2\,,
\end{equation}
and the leptonic invariant mass $q^2$ can be re-expressed in terms of partonic quantities
using (\ref{eq:q})
\begin{align}
\label{eq:conscheckMX}
    q^2 &= m_b^2 -2m_b \varepsilon_X + m_X^2\,,\nonumber\\
    M_X^2 &= \bLam^2 + 2 \bLam \varepsilon_X  + m_X^2\,.
\end{align}
It follows that the moments of $q^2$ and $M_X^2$  can be expressed in terms of the same building blocks, namely the moments
of the partonic energy and invariant mass. Their expressions are related by the replacement $\bLam \leftrightarrow -m_b$.   This simple kinematic relation allows for useful checks.

Following \cite{Alberti:2012dn} and assuming massless leptons, we write the triple differential width as 
\begin{equation}
\label{eq:d3Gamma}
\frac{d^3 \Gamma}{d\hat{q}^2 \,d\hat{u}\, d\hat{E}_\ell} =\Gamma_0\theta(\hat{u}_+ - \hat{u})\biggl\{\hat{q}^2 W_1 - \biggl[2\hat{E}_\ell^2 -2 \hat{E}_\ell \hat{q}_0 + \frac{\hat{q}^2}{2} \biggr]W_2 +\hat{q}^2 (2\hat{E}_\ell - \hat{q}_0) W_3 \biggr\}\,,
\end{equation}
where $\Gamma_0=|V_{cb}|^2 G_F^2 m_b^5/(16\pi^3)$,
the $W_i$ are structure functions which encode the hadronic physics and are functions  of $q^2$ and $u$, but not of $E_\ell$.
The structure functions admit  an OPE and can be written, in the on-shell scheme, as
\begin{align}
\label{eq:Wi}
W_i =& W_i^{(0,0)} + W_i^{(\pi, 0)}\frac{\mu_\pi^2}{2m_b^2} + W_i^{(G,0)}\frac{\mu_G^2}{2m_b^2} + W_i^{(D,0)}\frac{\rho^3_D}{2m_b^3} + W_i^{(LS,0)}\frac{\rho^3_{LS}}{2m_b^3}\nonumber\\
&+ \frac{\as(\mu_s) C_F}{\pi}\biggl[W_i^{(0,1)} + W_i^{(\pi, 1)}\frac{\mu_\pi^2}{2m_b^2} + W_i^{(G,1)}\frac{\mu_G^2}{2m_b^2} + W_i^{(D,1)}\frac{\rho^3_D}{2m_b^3} + W_i^{(LS,1)}\frac{\rho^3_{LS}}{2m_b^3} \biggr] \nonumber\\
&+\frac{\as^2 \beta_0 C_F}{\pi^2} \Bigl(W_i^{(0,2)}+ \frac{1}{2}\ln \frac{\mu_s}{m_b}W_i^{(0,1)}\Bigr) +O\Bigl(\as^2, \frac{\LamQCD^4}{m_b^4}\Bigr)\,,
\end{align}
where $\beta_0 = 11- \frac{2}{3}n_f $ and we have retained only the known contributions.
Throughout this work the coupling constant $\as \equiv \as^{(n_f=4)}(\mu_s)$ is evaluated at the generic scale $\mu_s$, and indicated explicitly in the terms where it makes a difference.
The nonperturbative parameters $\mu_\pi^2$, $\mu_G^2$, $\rho_D^3$, $\rho_{LS}^3$ are matrix elements of local operators in the {\it physical } $\bar{B}$ meson states (not taking the $m_b\to \infty$ limit), see \cite{Alberti:2013kxa,Gremm:1996df}. Furthermore, unless otherwise specified,
the $\overline{\rm MS}$ scale of $\mu_G^2$, $\rho_D^3$ and $\rho_{LS}^3$ is set to $m_b$.

Our goal is to compute the $\hat q^2$ spectrum from Eqs.~(\ref{eq:d3Gamma}, \ref{eq:Wi}).
Integrating over the lepton energy  gives
\begin{equation}
\label{eq:d2Gamma}
\frac{d^2 \Gamma}{d\hat{q}^2 \,d\hat{u}} =  \Gamma_0 \theta(\hat{u}_+ - \hat{u})\sqrt{\hat{q}_0^2 - \hat{q}^2}\biggl\{\hat{q}^2 W_1 +\frac{1}{3}(\hat{q}_0^2 - \hat{q}^2) W_2 \biggr\}\,,
\end{equation}
where $\hat{q}_0$ is to be seen as the function of $\hat{u}$ and $\hat{q}^2$ given below~\eqref{eq:q2lims}. We notice that the function $W_3$ drops out when integrating over the whole lepton energy range.

For the integration in $\hat{u}$ we need the explicit expressions for the structure functions, which can be found in~\cite{Manohar:1993qn,Blok:1993va,Gremm:1996df,Alberti:2012dn,Alberti:2013kxa,Colangelo:2020vhu}, except for $W_i^{(D,1)}$ and $W_i^{(LS,1)}$. 
Because of reparametrization invariance (RPI) the $O(\alpha_s \rho_{LS}^3/m_b^3)$ contribution to the $q^2$ spectrum  is proportional to the one of $O(\alpha_s \mu_G^2/m_b^2)$ \cite{Manohar:2010sf,Fael:2018vsp},
while the $O(\alpha_s\rho_D^3 /m_b^3)$ contribution to the $q^2$ spectrum is given in~\cite{Mannel:2021zzr} without recourse to the structure functions.

We write the result of integrating~\eqref{eq:d2Gamma} in $\hat{u}$ as
\begin{align}
\label{eq:Gamma}
\frac{d \Gamma}{d\hat{q}^2} &= \Gamma_0 \biggl[S^{(0,0)} + S^{(\pi, 0)}\frac{\mu_\pi^2}{2m_b^2} + S^{(G,0)}\frac{\mu_G^2}{2m_b^2} + S^{(D,0)}\frac{\rho^3_D}{2m_b^3} + S^{(LS,0)}\frac{\rho^3_{LS}}{2m_b^3}\nonumber\\
&+ \frac{\as(\mu_s) C_F}{\pi}\biggl(S^{(0,1)} + S^{(\pi, 1)}\frac{\mu_\pi^2}{2m_b^2} + S^{(G,1)}\frac{\mu_G^2}{2m_b^2} + S^{(D,1)}\frac{\rho^3_D}{2m_b^3} + S^{(LS,1)}\frac{\rho^3_{LS}}{2m_b^3} \biggr) \nonumber\\
&+\frac{\as^2 \beta_0 C_F}{\pi^2}\Bigl(S^{(0,2)} + \frac{1}{2}\ln \frac{\mu_s}{m_b}S^{(0,1)} \Bigr)+O\Bigl(\as^2, \frac{\LamQCD^4}{m_b^4}\Bigr)\biggr]\,,
\end{align}
where $S^{(p,n)}$ are functions of $\hat{q}^2$ with $p$ labelling the term in the heavy mass expansion and $n$ the perturbative order in $\as$.
Whenever the $W_i^{(p,n)}$ are available, they can be computed from 
\begin{equation}
S^{(p,n)}(\hat{q}^2) = 4\int_{-\infty}^{\infty} d\hat{u} \,\theta(\hat{u}_+ -\hat{u})\sqrt{\hat{q}_0^2-\hat{q}^2}\biggl\{3\hat{q}^2 W_1^{(p,n)}(\hat{q}^2,\hat{u}) +(\hat{q}_0^2 - \hat{q}^2) W_2^{(p,n)}(\hat{q}^2,\hat{u}) \biggr\}\,.
\end{equation}
The results are conveniently expressed in terms of 
\begin{equation}
    \omega= \frac{1}{2}(1+\hat{q}^2-\rho) = \hat{q}_0^{\rm max} \,.
\end{equation}
The leading order result is
\begin{equation}
    S^{(0,0)}(\hat{q}^2) = 8\sqrt{\omega^2 - \hat{q}^2} (\hat{q}^2-3\hat{q}^2 \omega +2\omega^2)\,,
\end{equation}
in agreement with \cite{Fael:2018vsp,Mannel:2021zzr}. 
The $O(1/m_b^2) $ power corrections  at leading order in $\as$ are
\begin{align}
    S^{(\pi,0)}(\hat{q}^2) &= -S^{(0,0)}(\hat{q}^2)\,,\\
    S^{(G,0)}(\hat{q}^2) &= \frac{8}{\sqrt{\omega ^2-\hat{q}^2}}\Bigl(6 \hat{q}^4 (2 \omega -1)+3 \hat{q}^2 \omega^2(1-5 \omega)  + 2\omega^3(5 \omega -2) \Bigr)\,,
\end{align}
 in agreement with RPI and \cite{Fael:2018vsp,Mannel:2021zzr}, respectively.
We  compute the leading order expressions for the $\rho^3_{LS}$ and $\rho^3_D$ terms from  
$W_i(\hat{q}^2,\hat{u}) = \text{Im}[T_i]/(2\pi)$ of \cite{Gremm:1996df}, or equivalently from the results of 
\cite{Colangelo:2020vhu} after using $\hat{\mu}^2_G = \mu_G^2 -\frac{\rho^3_{LS}+\rho^3_D}{m_b}$.
The results for the $q^2$ spectrum are 
\begin{align}
    S^{(LS,0)}(\hat{q}^2) &= -S^{(G,0)}(\hat{q}^2)\,,\nonumber\\
    S^{(D,0)}(\hat{q}^2) &= -\frac{8}{3 \left(\omega ^2-\hat{q}^2\right)^{3/2}}\Bigl[3 \hat{q}^6+3 \hat{q}^4 \left(-6 \omega^2+2 \omega +1\right) \omega \nonumber\\
    &\qquad\qquad\qquad\qquad+\hat{q}^2 \left(15 \omega ^2+7 \omega -22\right) \omega ^3+2 (8-5 \omega ) \omega
   ^5\Bigr] \,,
\end{align}
where $S^{(D,0)}$ agrees with $-\mathcal{C}^{(0)}_{\rho_D}/(24\pi^2)$ from~\cite{Mannel:2021zzr}.
Unlike $S^{(0,0)}$ and $S^{(G,0)}$, $S^{(D,0)}$ has a non-integrable singularity at the endpoint $\hat q^2_{\rm max}=(1-\sqrt{\rho})^2$,
\begin{equation}
    S^{(D,0)}_{\rm sing}(\hat{q}^2)=8 \frac{(1-\sqrt{\rho})^2 \rho^{\frac14}}{[1-\hat q^2/\hat q^2_{\rm max}]^{3/2}},
\end{equation}
which originates in the kinematic square root  in Eq.~(\ref{eq:d2Gamma}). The derivatives of this non-analytic term emerge from the integration of the $\delta^{(n)} $ that appear in the $W_i$ and are more and more singular at the endpoint for increasing $n$. Notice that the singularity appears only in the $q^2$ spectrum, and not in the total
rate or in the moments of the $q^2$ spectrum, which can be computed from the double differential distribution
(\ref{eq:d2Gamma}) using a different order of integration. It is therefore  possible to introduce a regularisation. Ref.~\cite{Mannel:2021zzr} employs dimensional regularisation, but other choices would lead to equivalent results. The regularised spectrum is then obtained by replacing the singular term with a plus distribution and a Dirac $\delta$:
\begin{equation}
    S^{(D,0)}_{\rm sing}(\hat{q}^2)\to 8 (1-\sqrt{\rho})^2 \rho^{\frac14}  \left(-2\, \delta(1-\xi) + 
    \left[\frac1{\left(1-\xi\right)^{3/2}}\right]_+
    \right),
\end{equation}
with $\xi=\hat q^2/\hat q^2_{\rm max}$. Here the plus distribution 
is defined in the usual way,
\begin{equation}
    \int_0^1  f(\xi) \left[\frac1{\left(1-\xi\right)^{3/2}}\right]_+ d\xi=
    \int_0^1 \frac{f(\xi)-f(1)}{\left(1-\xi\right)^{3/2}} d\xi,
\end{equation}
with $f(\xi)$ an arbitrary test function.

Let us now turn to  the $O(\as)$ contributions to (\ref{eq:Gamma}). Starting from the $O(\as)$ contributions to the form factors $W_i$ given in \cite{Aquila:2005hq,Alberti:2012dn}, we find excellent numerical agreement with \cite{Mannel:2021zzr}, namely
\begin{equation}
    S^{(0,1)}(\hat{q}^2) = \frac{\mathcal{C}_0^{\rm NLO,F}}{24\pi^2}\,,
\end{equation}
where $\mathcal{C}_0^{\rm NLO,F}$ is given in Eq.~(50) of \cite{Mannel:2021zzr}.

For the $O(\as\mu_\pi^2)$ term we have verified numerically that the formulas for $W_i^{(\pi,1)}$ in~\cite{Alberti:2012dn} lead to  the relation
\begin{equation}
    S^{(\pi,1)}(\hat{q}^2) = -S^{(0,1)}(\hat{q}^2)\,.
\end{equation}
 We proceed in the same way for the
 $O(\as\mu_G^2)$ term using $W_i^{(G,1)}$ from \cite{Alberti:2013kxa}. However, the non-perturbative parameters are defined differently in~\cite{Mannel:2021zzr} and~\cite{Alberti:2013kxa}, as the former includes the perturbative Wilson coefficient of the chromomagnetic operator in the definition of $\mu_G^2$,
\begin{equation}
\mu_G^2\Bigl|_{\text{\cite{Mannel:2021zzr}}} =    c_F(\mu)\, \mu_G^2(\mu) \Bigl|_{\text{\cite{Alberti:2013kxa}}} \,,
\end{equation}
where
\begin{equation}
    c_F(\mu) = 1 +\frac{\as}{2\pi}\biggl[C_F + C_A\biggl( \ln \frac{\mu}{m_b} + 1\biggr) \biggr]\,.
\end{equation}
Taking this into account we have verified that
\begin{equation}
    S^{(G,1)}(\hat{q}^2) = \frac{\mathcal{C}_{\mu_G}^{(1)} + \mathcal{C}_{\mu_G}^{(0)}c_F^{(1)}(m_b)}{24\pi^2}\,.
\end{equation}
where $\mathcal{C}^{(0,1)}_{\mu_G}$ are given in \cite{Mannel:2021zzr}.

Finally, 
the $O(\as \rho_D^3)$ terms have only recently been computed in~\cite{Mannel:2021zzr} and we employ their results
\begin{align}
    S^{(D,1)}(\hat{q}^2) &= - \frac{\mathcal{C}_{\rho_D}^{(1)} + c_D^{(1)}(m_b)\mathcal{C}_{\rho_D}^{(0)}}{24\pi^2} \,,
\end{align}
where we have again taken into account that\footnote{Notice that $\rho_D^3$ defined as in~\cite{Mannel:2021zzr} is still  $\mu$-dependent because it mixes with other dimension-6 operators under renormalization.} 
\begin{equation}
\rho_D^3(\mu)\Bigl|_{\text{\cite{Mannel:2021zzr}}} =    c_D(\mu) \rho_D^3(\mu)\Bigl|_{\text{\cite{Alberti:2013kxa}}} \,,
\end{equation}
with the Darwin operator Wilson coefficient in the HQET Lagrangian given by
\begin{equation}
    c_D(\mu) = 1 + \frac{\as C_F}{\pi}\biggl[-\frac{8}{3} \ln \frac{\mu}{m_b} + \frac{C_A}{C_F}\biggl(\frac{1}{2}-\frac{2}{3}\ln \frac{\mu}{m_b} \biggr) \biggr]\,.
\end{equation}
As previously stated the $O(\as \rho_{LS}^3)$ corrections in the on-shell scheme can be inferred from reparametrization invariance\footnote{This RPI relation holds only if the Wilson coefficients of the HQET Lagrangian are \emph{not} included in the definition of the non-perturbative parameters,  in conflict with eq.(39) of~\cite{Mannel:2021zzr}.}~\cite{Manohar:2010sf}
\begin{equation}
    S^{(LS,1)}(\hat{q}^2) = -S^{(G,1)}(\hat{q}^2)\,.
\end{equation}

\begin{figure}
    \centering
    \includegraphics[width=0.7\textwidth]{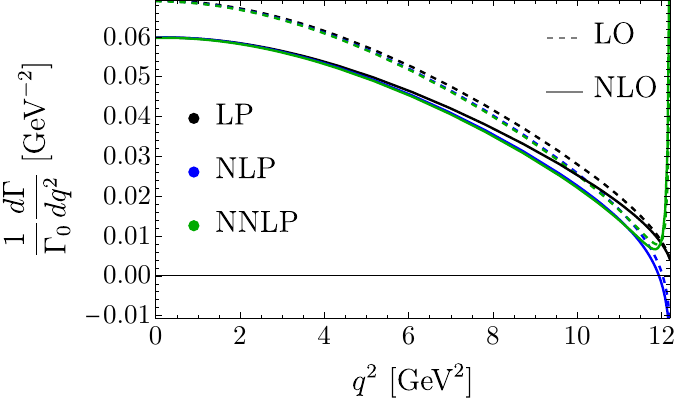}
    \caption{The $q^2$ spectrum computed in the on-shell scheme at different orders in the perturbative expansion (dashed LO and solid for NLO) and in the heavy quark expansion with LP, NLP (up to $O(1/m_b^2)$) and NNLP (up to $O(1/m_b^3)$) respectively in black, blue and green. We used the  inputs $m_b=4.8~\text{GeV}$, $\rho=0.073$, $\mu_\pi^2=0.3~\text{GeV}^2$, $\mu_G^2=0.35~\text{GeV}^2$, $\rho_D^3=0.1~\text{GeV}^3$, $\rho_{LS}^3=-0.15~\text{GeV}^3$ and $\as(m_b) =0.219$.}
    \label{fig:spect}
\end{figure}

In Figure~\ref{fig:spect} we plot the $\hat{q}^2$ spectrum as a function of $q^2$, 
normalized to the standard prefactor $\Gamma_0$. 
The spectrum is shown  up to $q^2 = 12.2~\rm{GeV}$ and one can see the rise of the endpoint 
singularity related to the $O(\rho_D^3)$ contribution at $q^2_{\rm max} = (m_b-m_c)^2=12.27~{\rm GeV}^2$.
We note that  the perturbative corrections to the spectrum are sizeable, but they largely cancel in the normalised moments, where the power corrections will be more important as they significantly change the spectrum at high $q^2$.

\subsection{The moments}
We define the $q^2$ moments as
\begin{equation}
M_n(q^2_{\rm cut}) = m_b^{2n} \int_{\hat{q}^2_{\rm cut}}^{\hat{q}^2_{\max}} d\hat{q}^2 \frac{d\Gamma}{d\hat{q}^2} \hat{q}^{2n}\,,
\end{equation}
and the normalized moments as
\begin{equation}
\langle q^{2n}\rangle
%\equiv \hat{M}_n(q^2_{\rm cut}) 
= \frac{M_n(q^2_{\rm cut})}{M_0(q^2_{\rm cut})}\,.
\end{equation}
In the above ratios we always re-expand the moments in $\as$ and $1/m_b$.
We study the first three central moments,
\begin{equation}
    Q_1= \langle q^2\rangle, \qquad Q_2=\langle (q^2 -\langle q^2 \rangle)^2\rangle,\qquad Q_3=\langle (q^2 -\langle q^2 \rangle)^3\rangle,
\end{equation}
for different values of $q^2_{\rm cut}$.
We start with the results in the on-shell scheme. After expanding in $\as$ and in $1/m_b$ and neglecting higher order terms,  each moment $Q_i$ can be expressed as 
\begin{align}
\label{eq:onshell}
    Q_i =& \tilde{Q}_i^{(0,0)} + \biggl(\frac{\mu^2_G}{2m_b^2}-\frac{\rho^3_{LS}}{2m_b^3}\biggr)\tilde{Q}_i^{(G,0)} +\frac{\rho^3_D}{2m_b^3}\tilde{Q}_i^{(D,0)} \nonumber\\
    &+ \frac{\as(\mu_s) C_F}{\pi} \biggl(\tilde{Q}_i^{(0,1)} + \biggl(\frac{\mu^2_G}{2m_b^2}-\frac{\rho^3_{LS}}{2m_b^3}\biggr) \tilde{Q}_i^{(G,1)} +\frac{\rho^3_D}{2m_b^3}\tilde{Q}_i^{(D,1)} \biggr) \nonumber\\
    &+ \frac{\as^2 C_F}{\pi^2}\beta_0 \Bigl(\tilde{Q}_i^{(0,2)} + \frac{1}{2}\ln \frac{\mu_s}{m_b}\tilde{Q}_i^{(0,1)} \Bigr)\,,
\end{align}
where the $\mu_\pi^2$ terms have dropped out. 
Again, the renormalization scale of $\mu_G^2$,  $\rho_D^3$ and $\rho^3_{LS}$ is set equal to $m_b$.
Table \ref{tab:onshell} presents the results of the various contributions in (\ref{eq:onshell}) for two representative values of $q^2_{\rm cut}$, using  $m_b=4.8~\rm{GeV}$ and $\rho = 0.073$.  The leading power perturbative corrections are invariably small, while the power corrections are sizeable, especially for the higher moments.

\begin{table}[t]
\vspace{4mm}
\centering
\begin{tabular}{|c|c|c|c|c|c|c|c|}
\hline
$q^2_{\rm cut} $ & $\tilde{Q}_1^{(0,0)}$  & $\tilde{Q}_1^{(0,1)}$ & $\tilde{Q}_1^{(0,2)}$  & $\tilde{Q}_1^{(G,0)}$ & $\tilde{Q}_1^{(G,1)}$ & $\tilde{Q}_1^{(D,0)}$ & $\tilde{Q}_1^{(D,1)}$\\
\hline
2 & $5.9731$ & $0.34754$ & $0.17783$ & $-25.268$ & $-61.525$ & $-213.29$ & $-662.51$\\
8 & $9.6494$ & $0.13734$ & $0.18177$ & $-27.423$ & $-67.705$ & $-381.96$ & $-1078.0$\\
\hline
$q^2_{\rm cut}$ & $\tilde{Q}_2^{(0,0)}$  & $\tilde{Q}_2^{(0,1)}$ & $\tilde{Q}_2^{(0,2)}$  & $\tilde{Q}_2^{(G,0)}$ & $\tilde{Q}_2^{(G,1)}$ & $\tilde{Q}_2^{(D,0)}$ & $\tilde{Q}_2^{(D,1)}$\\
\hline
2 & $6.9027$ & $0.66577$ & $0.71576$ & $-94.050$ & $-231.92$ & $-1179.6$ & $-3488.5$\\
8 & $1.2201$ & $0.091747$ & $0.15975$ & $-41.946$ & $-104.21$ & $-999.68$ & $-2723.4$\\
\hline
$q^2_{\rm cut}$ & $\tilde{Q}_3^{(0,0)}$  & $\tilde{Q}_3^{(0,1)}$ & $\tilde{Q}_3^{(0,2)}$ &$\tilde{Q}_3^{(G,0)}$ & $\tilde{Q}_3^{(G,1)}$ & $\tilde{Q}_3^{(D,0)}$ & $\tilde{Q}_3^{(D,1)}$ \\
\hline
2 & $7.0650$ & $-1.1411$ & $0.80766$ & $-170.52$ & $-395.81$ & $-4732.2$ & $-12760.$\\
8 & $0.53627$ & $-0.11087$ & $-0.097089$ & $-29.736$ & $-67.629$ & $-1888.2$ & $-4950.1$\\
\hline
\end{tabular}
\caption{Values of the first three central moments of the $q^2$ spectrum in the on-shell scheme for two values of $q^2_{\rm cut}$. All quantities are in GeV to the appropriate power. }
\label{tab:onshell}
\end{table}

\subsection{BLM corrections}
The $O(\as^2 \beta_0)$ corrections to the hadronic structure functions and therefore to the triple differential distribution have been computed in Ref.~\cite{Aquila:2005hq}, which gives  explicit numerical results for the leptonic and hadronic form factors,
as well as a  \texttt{FORTRAN} code implementing the calculation. 
This code performs a multidimensional numerical integration over the gluon mass, the variable $u$, two Feynman parameters and $q^2$, and can compute directly the BLM corrections to the $q^2$ moments. However, 
due to very large cancellations between the different terms, reaching a good precision in the correction to $\tilde{Q}_i^{(2,0)}$ is a daunting task, especially for $i=2,3$ (the same problem affects the lepton energy moments).
We have therefore first computed with high precision $S^{(0,2)}(q^2)$, namely the BLM correction to the spectrum~\eqref{eq:Gamma}, from which we then compute the corrections to the moments at different values of $q^2_{\rm cut}$ by a simple one dimensional integration. This procedure allows us to reach a much higher numerical accuracy, especially for $\tilde{Q}_2^{(2,0)}$ and $\tilde{Q}_3^{(2,0)}$.

The function $S^{(0,2)}(q^2)$ is obtained by a mix of analytical and numerical integrations at 60 values of $q^2$ in the whole allowed range.
The set of points is then interpolated with a cubic polynomial of coefficients $c_i$, reproducing 
$S^{(0,2)}(q^2)$ with excellent accuracy. 
To include the $\rho$ dependence, we repeated this procedure for 6 different values of $\rho \in [0.035, 0.087]$.
The coefficients of the cubic polynomial are then turned into $c_i(\rho)$ by another cubic fit to the 6 points in $\rho$.
This allows us to have an interpolated formula for $S^{(0,2)}(q^2,\rho)$, and the quality of the interpolation is checked by comparing the moments with the direct numerical integration. In the case of $\tilde{Q}_{2,3}^{(2,0)}$ the uncertainty  is much smaller than the one we  have combining the results of numerical integration for the linear moments $M_i$.
In the supplemental material we provide approximate formulas for $\tilde{Q}_i^{(2,0)}$ for values of $\rho$ in the above range.

\subsection{Change to the kinetic scheme and results}
We are now ready  to translate our predictions to the kinetic scheme \cite{Bigi:1996si}. Since we do not have the complete
$O(\as^2)$ contributions to the $q^2$-moments, we perform the change of scheme including  only
the $O(\as^2 \beta_0)$ terms, given in \cite{Czarnecki:1997sz}. Notice that 
the three-loop relations between on-shell and kinetic scheme parameters have been computed in \cite{Fael:2020iea}. Denoting by $m_b(\mu_k) $
the kinetic mass at the cutoff scale $\mu_k$, the on-shell $b$ mass is given by its $\mu_k\to 0$ limit, and similarly for $\mu_\pi^2$
and $\rho_D^3$.
The relevant relations are
\begin{align}
\label{eq:OStoKin}
    m_b &\equiv m_b(0) = m_b(\mu_k) + [\bar{\Lambda}(\mu_k)]_{\rm pert} + \frac{[\mu^2_\pi(\mu_k)]_{\rm pert}}{2m_b(\mu_k)}\,,\nonumber\\
    \mu^2_\pi(0) &= \mu^2_\pi(\mu_k) - [\mu^2_\pi(\mu_k)]_{\rm pert}\,, \qquad \rho^3_D(0) = \rho^3_D(\mu_k) - [\rho^3_D(\mu_k)]_{\rm pert}\,,
\end{align}
with
\begin{align}
    [\bar{\Lambda}(\mu_k)]_{\rm pert} &= \frac{4}{3}\frac{\as(m_b) C_F}{\pi} \mu_k \biggl[1+ \frac{\as}{\pi}\biggl(\frac{\beta_0}{2}\Bigl(\ln \frac{m_b}{2\mu_k} + \frac{8}{3} \Bigr) -C_A \Bigl(\frac{\pi^2}{6} -\frac{13}{12} \Bigr)\biggr) \biggr]\,,\nonumber\\
    [\mu_\pi^2(\mu_k)]_{\rm pert} &= \frac{3}{4}\mu_k [\bar{\Lambda}(\mu_k)]_{\rm pert} -\frac{\as^2 C_F}{\pi^2}\beta_0 \frac{\mu_k^2}{4}\,,\nonumber\\
    [\rho^3_D(\mu_k)]_{\rm pert} &= \frac{\mu_k^2}{2}[\bar{\Lambda}(\mu_k)]_{\rm pert} -\frac{\as^2 C_F}{\pi^2}\beta_0 \frac{2\mu_k^3}{9}\,,
\end{align}
where as stated above we neglect the term proportional to $C_A$ in $[\bar{\Lambda}(\mu_k)]_{\rm pert}$, since it is not enhanced by $\beta_0$.
We will choose as our default value $\mu_k =1~\rm{GeV}$.

For what concerns the charm mass, we adopt the $\overline{\rm MS}$ scheme at a scale $\mu_c$, choosing as a reference $\mu_c=2$ GeV. The well-known relation between the pole and the  $\overline{\rm MS}$ charm mass \cite{Melnikov:2000qh} is
\begin{align}
    m_c=& \overline{m}_c(\mu_c)\biggl[1+\frac{\as(m_b)C_F}{\pi} \left(\frac{3}{2} \ln\frac{\mu_c}{\overline{m}_c}+1\right)\\
    &+\frac{\as^2 C_F}{\pi^2}\beta_0\biggl(\frac{3}{8} \ln^2 \frac{\mu_c}{\overline{m}_c} + \frac{13}{16} \ln \frac{\mu_c}{\overline{m}_c} + \frac{\pi^2}{16}+\frac{71}{128} + \frac{1}{2}\ln \frac{m_b}{\mu_c}\left(\frac{3}{2} \ln\frac{\mu_c}{\overline{m}_c}+1\right) \biggr) \biggr]\,. \nonumber
\end{align}
Before we present numerical results in the kinetic scheme we recall that since the kinetic scheme employs the  hard cutoff
$\mu_k$ in the $b$ quark reference frame, it breaks Lorentz invariance and RPI.\footnote{After a change of scheme for $m_b$ and the HQE parameters,  RPI still implies definite relations between the Wilson coefficients of the HQE parameters. However, these relations change order by order, unlike in the native HQET (on-shell) derivation; this is what we mean by breaking RPI. 
However, if one employs RPI to reduce the number of HQE parameters as in Refs. \cite{Fael:2018vsp,Bernlochner:2022ucr},
the perturbative scheme must preserve the RPI identities otherwise the reduction of parameters may be violated by perturbative corrections. }
For instance, in the kinetic scheme the two terms in the combination
\begin{equation}
  \frac{\mu_G^2}{2m_b^2} - \frac{\rho^3_{LS}}{2m_b^3}\,,
\end{equation}
that multiplies $\tilde Q_i^{(G,j)}$ in (\ref{eq:onshell}) receive different $O(\alpha_s)$ corrections due to the different power of $m_b$ in the denominator.
Therefore in the kinetic scheme we write the analogue of (\ref{eq:onshell}) as 
\begin{align}
    Q_i =& ~Q_i^{(0,0)} + \Big(\frac{\mu^2_G}{2m_b^2}-  \frac{\rho^3_{LS}}{2m_b^3}\Big)  Q_i^{(G,0)}
    %-\frac{\rho^3_{LS}}{2m_b^3}Q_i^{(LS,0)} 
    +\frac{\rho^3_D}{2m_b^3}Q_i^{(D,0)} \nonumber\\
    &+ \frac{\as(\mu_s) C_F}{\pi} \biggl(Q_i^{(0,1)} + \frac{\mu^2_G}{2m_b^2}Q_i^{(G,1)}-\frac{\rho^3_{LS}}{2m_b^3} Q_i^{(LS,1)} +\frac{\rho^3_D}{2m_b^3}Q_i^{(D,1)} \biggr) \nonumber\\
    &+ \frac{\as^2 C_F}{\pi^2} \beta_0 \Bigl(Q_i^{(0,2)} + \frac{1}{2}\ln \frac{\mu_s}{m_b}Q_i^{(0,1)}\Bigr)\,,
\end{align}
where $Q_i^{(G,1)}$ differs from $Q_i^{(LS,1)}$. Table~\ref{tab:momtabk} reports the coefficients of this expansion
for $q^2_{\rm cut}=2$ and 8 GeV$^2$ using $\mu_k=1~$GeV and the central values $m_b(1~\rm GeV)=4.573$ GeV and $\overline{m}_c(2~\rm GeV)=1.092$ GeV from the default fit of \cite{Bordone:2021oof}.
\begin{table}
\centering
\begin{tabular}{|c|c|c|c|c|c|c|c|c|}
\hline
$q^2_{\rm cut} $ & $Q_1^{(0,0)}$  & $Q_1^{(0,1)}$ & $Q_1^{(0,2)}$  & $Q_1^{(G,0)}$ & $Q_1^{(G,1)}$ & $Q_1^{(LS,1)}$ & $Q_1^{(D,0)}$ & $Q_1^{(D,1)}$\\
\hline
2 & $5.8535$ & $0.23657$ & $1.3695$ & $-24.373$ & $-44.243$ & $-36.554$ & $-225.61$ & $-245.30$\\
8 & $9.5732$ & $0.16033$ & $1.6661$ & $-27.135$ & $-43.569$ & $-35.008$ & $-416.25$ & $-455.03$\\
\hline
$q^2_{\rm cut}$ & $Q_2^{(0,0)}$  & $Q_2^{(0,1)}$ & $Q_2^{(0,2)}$  & $Q_2^{(G,0)}$ & $Q_2^{(G,1)}$ & $Q_2^{(LS,1)}$ & $Q_2^{(D,0)}$ & $Q_2^{(D,1)}$\\
\hline
2 & $6.5996$ & $0.54190$ & $5.4250$ & $-90.767$ & $-110.90$ & $-82.269$ & $-1228.3$ & $-628.79$\\
8 & $1.1248$ & $1.4605$ & $4.2301$ & $-40.463$ & $-14.864$  & $-2.0994$ & $-1046.5$ & $-69.680$\\
\hline
$q^2_{\rm cut}$ & $Q_3^{(0,0)}$  & $Q_3^{(0,1)}$ & $Q_3^{(0,2)}$ &$Q_3^{(G,0)}$ & $Q_3^{(G,1)}$ & $Q_3^{(LS,1)}$ & $Q_3^{(D,0)}$ & $Q_3^{(D,1)}$ \\
\hline
2 & $7.0034$ & $3.9048$ & $16.720$ & $-174.73$ & $48.712$  & $103.84$ & $-4913.7$ & $1739.8$\\
8 & $0.49516$ & $4.7470$ & $8.2651$ & $-29.146$ & $51.317$ & $60.512$ & $-1917.5$ & $2205.8$\\
\hline
\end{tabular}
\caption{Table of numerical values for the expansion coefficients  of the central moments in the \emph{kinetic} scheme 
using $\mu_k=1~$GeV, $m_b(1~\rm GeV)=4.573$ GeV, and $\overline{m}_c(2~\rm GeV)=1.092$ GeV. All quantities are in GeV to the appropriate power.}
\label{tab:momtabk}
\end{table}

\begin{figure}
    \centering
    \includegraphics[width=\textwidth]{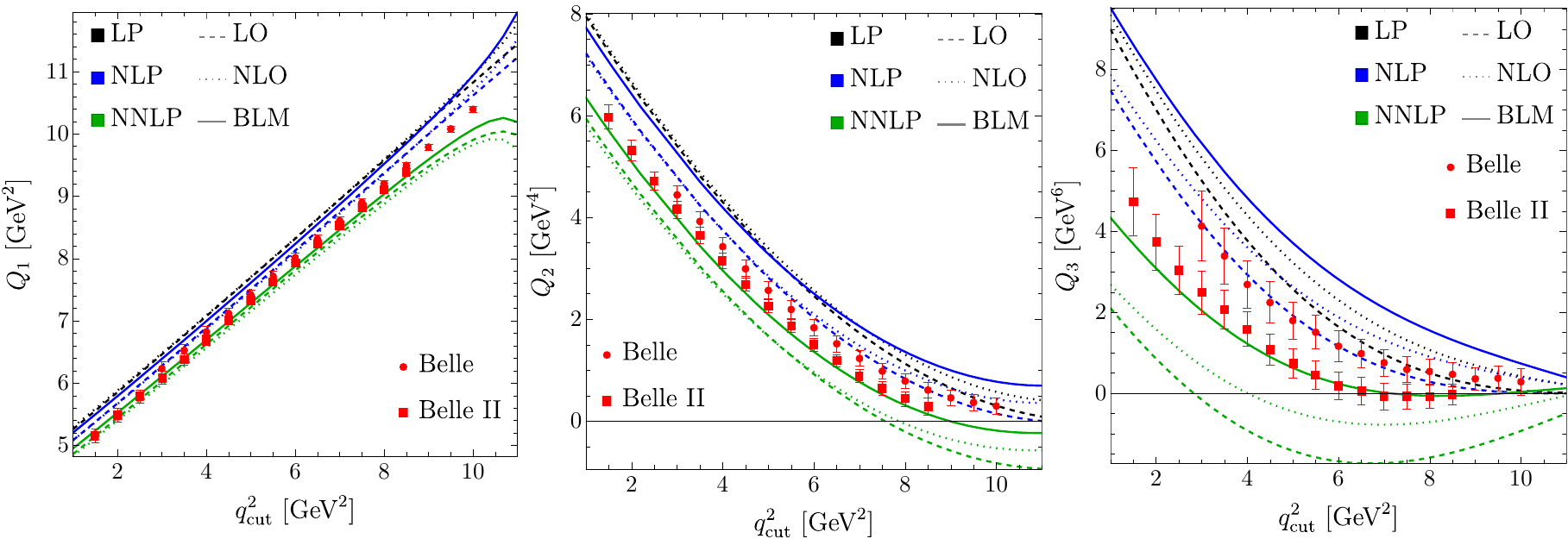}
    \caption{\small Comparison of the first three central moments in the \emph{kinetic} scheme between theoretical prediction and experimental data from Belle~\cite{Belle:2021idw} (red dots) and Belle II~\cite{Belle-II:2022fug} (red squares).
    The various curves represent calculations including all terms at leading power in $m_b$ (LP), up to $O(1/m_b^2)$ (NLP), up to $O(1/m_b^3)$ (NNLP), and up to $O(\as^0,\as^1, \as^2 \beta_0)$ (LO, NLO, BLM).
    }
    \label{fig:q2Momk}
\end{figure}

In Figure~\ref{fig:q2Momk} we show our results for the first three central $q^2$-moments as functions
of the minimum value of $q^2$, $q^2_{\rm cut}$. For the quark masses and  OPE parameters we have employed 
the central values and scale settings of the default fit in~\cite{Bordone:2021oof}. The results are compared with the Belle and Belle II measurements. We observe that the perturbative and power corrections to the first moment are small, except for 
very high $q^2_{\rm cut}$. On the other hand, the power corrections to the second and third central moments are quite sizeable, as expected.
We also observe that our calculation of the  variance $Q_2$, which is a positive  definite quantity,  becomes negative for values of $q^2_{\rm cut}$ slightly larger than 8 GeV$^2$. This is suggestive of relevant  higher order effects
in the high $q^2$ tail that are not included in our calculation. In our analysis of the next section we will not consider experimental 
data with $q^2_{\rm cut}> 8$ GeV$^2$.

\begin{figure}
    \centering
    \includegraphics[width=\textwidth]{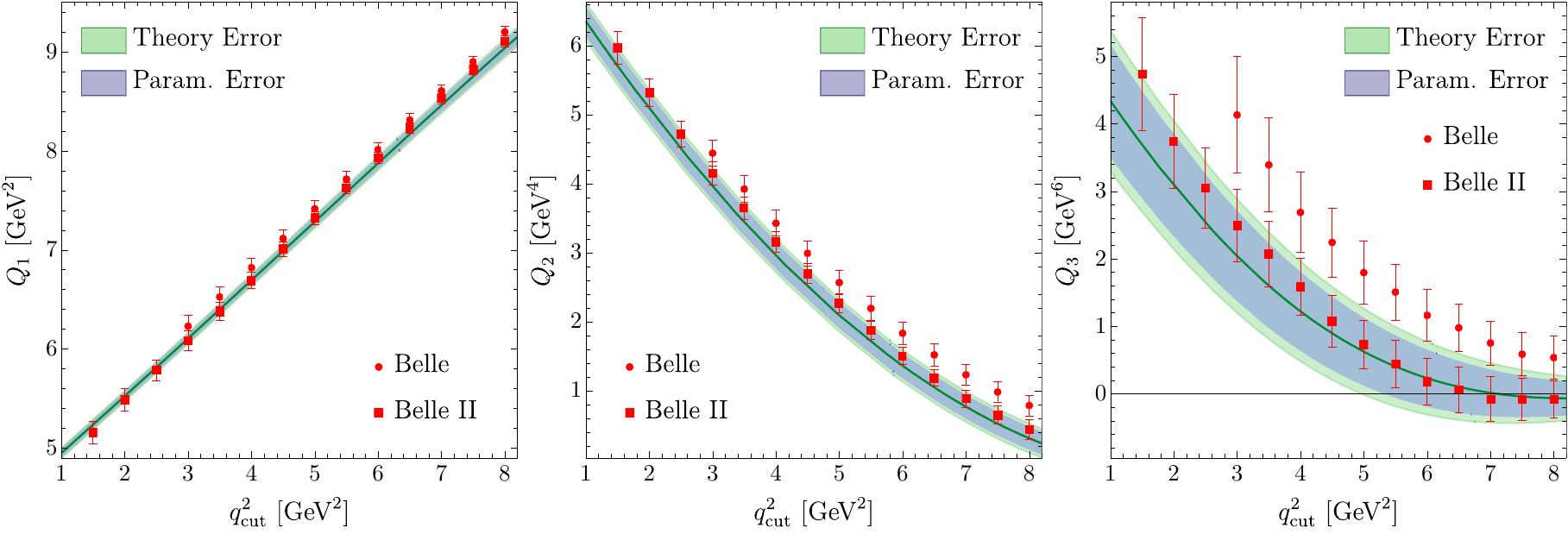}
    \caption{\small Results for the first three central moments including the theory uncertainty bands (green) and the parametric uncertainty from the fit~\cite{Bordone:2021oof} results (blue). The combined errors are not shown.}
    \label{fig:ploterrors}
\end{figure}

In Fig.~\ref{fig:q2Momk} we observe a good agreement between our predictions for the first $q^2$-moment and the Belle and Belle II measurements. However, the Belle measurements of $Q_2$ and $Q_3 $ are consistently above our predictions. In order to understand 
whether there is a serious  discrepancy, we show in Figure~\ref{fig:ploterrors} our predictions together with our estimate of the theory uncertainty and with the parametric uncertainty from the default fit of \cite{Bordone:2021oof}. To estimate the theory uncertainty we employ the same method as  \cite{Gambino:2013rza,Bordone:2021oof}, combining in quadrature the variations in $Q_i$ obtained varying 
the parameters by fixed amounts. The resulting uncertainties tend to be relatively conservative and improve 
significantly the agreement with the Belle and Belle II data. We note that the predictions of $Q_{2,3}$ are
particularly sensitive to $\rho^3_D$, and we can definitely expect the experimental data for $Q_{2,3}$ to have an important impact in the determination of $\rho^3_D$ through a global fit. 

While we will investigate this in the next section, we can already determine the size of $\rho^3_D$ preferred by the central values of the Belle and Belle II $Q_{2,3}$ measurements, assuming all other inputs are unchanged. In the case  $q^2_{\rm cut} =6$ GeV$^2$, the Belle and Belle II central values of $Q_3$ prefer $\rho^3_D\approx 0.12$ GeV$^3$ and 0.19 GeV$^3$, respectively, with an experimental uncertainty of around 0.03 GeV$^3$. Different values of $q^2_{\rm cut}$ lead to roughly similar results, with lower values of $\rho^3_D$ preferred (with larger experimental uncertainty) at lower $q^2_{\rm cut}$. Similarly, for $q^2_{\rm cut} =6$ GeV$^2$, the Belle and Belle II central values of $Q_2$  prefer $\rho^3_D\approx 0.11$ GeV$^3$ and 0.16 GeV$^3$, respectively, with an experimental uncertainty between 0.020 and 0.025 GeV$^3$. 
In summary, even considering the theory uncertainty of our predictions, 
the Belle data for $Q_{2,3} $ appear in tension with the results of the fit of \cite{Bordone:2021oof},
but they are also in tension with the Belle II results: for instance
$Q_3$ measured at $q^2_{\rm cut} =6$ GeV$^2$ by Belle and Belle II is  0.18(35) GeV$^6$ and 1.16(38) GeV$^6$, respectively (a $\sim 2\sigma $ tension). 
It is also worth mentioning that even the low range of  $\rho^3_D$ favoured by 
the Belle $q^2$-moments data is quite far from the results of the fit without 
higher power corrections in \cite{Bernlochner:2022ucr}.

The above considerations on $Q_{2,3} $ depend significantly on the inclusion of the BLM corrections in our predictions. Indeed, we see in Fig. \ref{fig:q2Momk} that they shift $Q_{2,3} $ up by an amount similar to the difference between then Belle and Belle II results. This is directly related to the large BLM contribution to the perturbative definition of $\rho^3_D(\mu)$, cfr. Eq.~(\ref{eq:OStoKin}). We also observe a sizeable residual
dependence on the scale $\mu_s$ at which $\as $ is evaluated (in Figs.~\ref{fig:q2Momk} and \ref{fig:ploterrors} we employ $\mu_s=m_b/2$), which  is however within our 
 theory uncertainty estimate.

The $O(\alpha_s^2) $ non-BLM corrections to the $q^2$ moments are only available for  $q^2_{\rm cut}=0$ \cite{Fael:2022frj}. 
Their size, relative to the BLM ones,   depends on the scheme and the scales employed for the masses. 
 It is useful to recall the case of the leptonic and hadronic moments:
 when both the $b$ and $c$ masses are in the on-shell or kinetic scheme the two-loop non-BLM corrections to the leptonic and hadronic moments are dominated by the BLM ones \cite{Gambino:2011cq}. 
%This applies to the total width as well as to the leptonic and hadronic moments, with the possibile exception of the second and third central moments, which were not known with a sufficient accuracy. 
However, the non-BLM corrections become larger when $m_c$ is expressed in the $\overline{\rm MS}$ scheme, especially for  $\mu_c= 3$~GeV. This is due to the large mass anomalous dimension, which enhances  non-BLM logarithms
in the calculation. As a consequence, the perturbative series tend to converge more slowly at $\mu_c= 3$~GeV and one has larger 
non-BLM contributions. This is the reason why we prefer a lower scale $\mu_c\sim 2$~GeV, for which the BLM corrections still 
provide the bulk of the complete $O(\as^2)$ corrections to leptonic and hadronic moments  \cite{Gambino:2011cq}.
Comparing our BLM calculation of the $q^2$ moments at $q^2_{\rm cut}=0$ with the calculation of Ref.~\cite{Fael:2022frj} we find that 
in the on-shell scheme the complete $O(\alpha_s^2)$ corrections to the first two central $q^2$ moments 
agree with the BLM result within 30\% or better. The non-BLM correction to the third central moment are 45\% of the BLM ones.
In the case of the kinetic scheme with $\overline{\rm MS}$ $c$ quark mass, Ref.~\cite{Fael:2022frj}  provides results only for $\mu_c=3$ GeV, Eq. (33), in which case  the non-BLM corrections are in general larger than the BLM ones.
We have also compared the results given in Eq.~(33)  at $O(\alpha_s^2)$ with our BLM calculation using $\mu_c=2$ GeV and $\mu_s=m_b/2$ (the default values used in the next Section) and find good numerical agreement and  deviations  between 20\% and 47\% of the BLM corrections. This suggests both the usefulness of the BLM computation (for small $\mu_c$) and the need for a complete $O(\alpha_s^2)$ calculation with generic cuts on $q^2$. 
 
\section{A global fit}
\label{sec:fit}
In this section we present fits to the  semileptonic moments that include the $q^2$ moments measured by Belle and Belle II.
As a first step, we repeat the  fit to hadronic and leptonic moments of Ref.~\cite{Bordone:2021oof} using exactly the same default inputs and settings, but including the new data. 
This will allow us to evaluate the impact of the new data on the global fit. Later on we will update some of the inputs, introduce a few improvements, and provide our final results.

\subsection{Impact of the $q^2$ moments}
In order to proceed with  the first step, we need to briefly review the strategy for the correlation among theoretical uncertainties 
adopted in \cite{Bordone:2021oof} and generalise  it to the case of the $q^2$ moments. Indeed, the treatment 
of central $q^2$-moments data measured with different lower cuts on $q^2$ presents problems analogous to those discussed 
in \cite{Gambino:2013rza} for the hadronic and leptonic moments measured with different lower cuts on the lepton energy. The extent to which the theoretical uncertainties of different observables are correlated is a subtle issue. In general, such  correlations are neglected but there are cases where this is unreasonable.
For instance, linear moments of different orders are certainly highly correlated, and the same holds for moments measured at nearby values of $q^2_{\rm cut}$.
In the default approach  employed in \cite{Bordone:2021oof} (option {\bf D} of \cite{Gambino:2013rza}) the theoretical uncertainties to different central moments are considered uncorrelated  and the correlation between the same moment at  different cuts on the lepton energy are modeled in such a way that the correlation is highest for adjacent cuts and low for distant cuts, and is lower  when the cuts get closer to the endpoint, where one expects higher order corrections to be more important. To extend this approach to the $q^2$-moments, we introduce a factor
\be 
\xi(q^2_{\rm cut})=1-\frac12 e^{-(\bar{q}^2- q^2_{\rm cut})/\Delta_q}\,,
\label{eq:xi}
\ee
which represents the correlation between  moments measured at $q^2$ cuts differing by  0.5 GeV$^2$.
The correlation between moments measured at $q^2$ cuts further away is given by the product
of $\xi$ computed at all intermediate values of $q^2_{\rm cut}$ spaced by 0.5 GeV$^2$. The parameters $\bar{q}^2$
and $\Delta_q$ are chosen in such a way that the correlation between far-away cuts and between 
nearby cuts 
close to the endpoint becomes small. Our default values are $\bar{q}^2=9$ GeV$^2$ and $\Delta_q=1.4$ GeV$^2$ and we will discuss later the effect of changing these values. 
As an illustration, the correlation between the theoretical errors of a generic moment with
cuts at 2 GeV$^2$ and 6 GeV$^2$    is
given by $\prod_{k=0}^7 \xi(2.25\;{\rm GeV}^2+ 0.5 k)=0.855$, while for cuts at 2 GeV$^2$ and 8 GeV$^2$ 
the correlation decreases to 0.49.

\begin{figure}
    \centering
    \includegraphics[width=\textwidth]{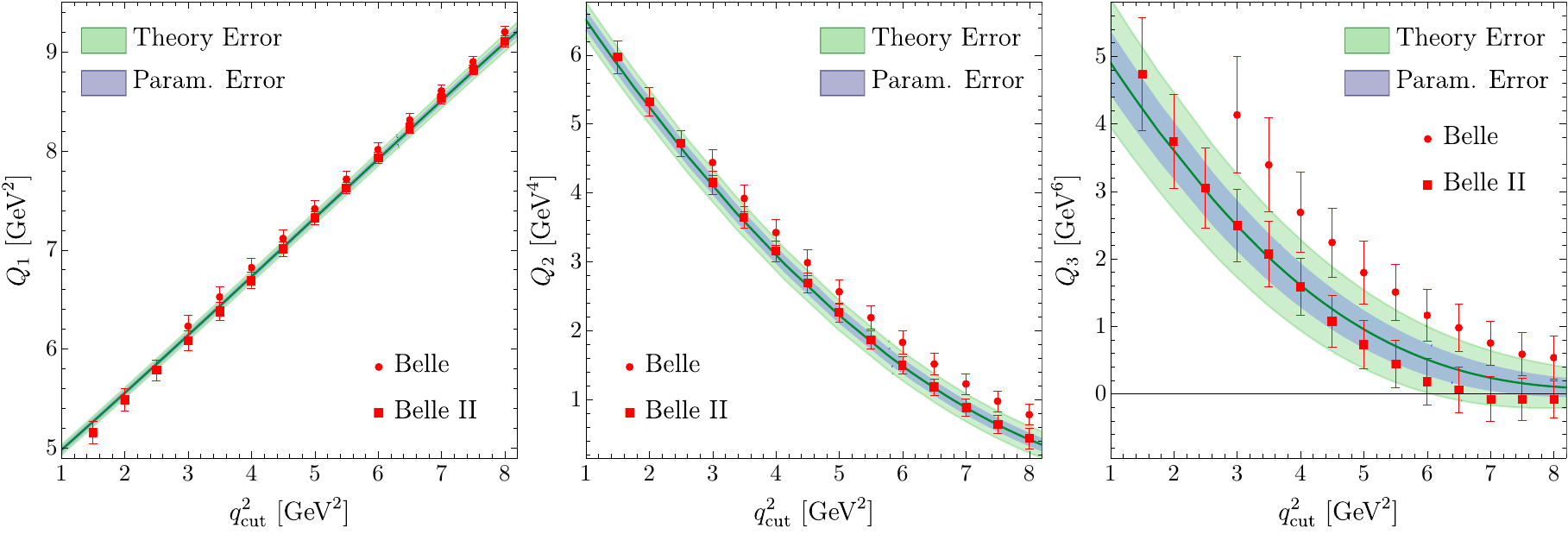}
    \caption{\small Results for the central moments including the theory uncertainty bands (green) and the parametric uncertainty from the results of the fit performed in this paper (blue). The combined errors are not shown.}
    \label{fig:ploterrorsq2}
\end{figure}

We compare the results of the default fit of \cite{Bordone:2021oof} with fits including the Belle and Belle II data in Table~\ref{tab:fitBelleBelleII}. 
Because of the large number of highly correlated data points and in analogy with the leptonic and hadronic moments, we
make a selection of the $q^2$-moments data:
from the Belle II dataset we choose the first three moments at five values of the lower cut $q^2_{\rm cut} = \{1.5, 3.0, 4.5, 6.0, 7.5\}~\text{GeV}^2$.
Similarly, in the Belle dataset we select the moments with $q^2_{\rm cut} = \{3.0, 4.5, 6.0, 7.5\}~\text{GeV}^2$. We have checked that the fits are very stable with respect to the 
choice of the subset of cuts to be included. We use the correlations between Belle and Belle II data that were employed in \cite{Bernlochner:2022ucr}.\footnote{We are grateful to the authors of \cite{Bernlochner:2022ucr} for sharing their covariance matrices for the $q^2$-moments.}
We see in Table~\ref{tab:fitBelleBelleII} that 
there is excellent agreement among the various fits, with a small downward shift of $\mu_\pi^2$ and $\rho^3_D$ (and consequently of $V_{cb}$) with respect to the results of~\cite{Bordone:2021oof}.
The uncertainty on $\rho^3_D$ is reduced significantly, but this reflects in only a small reduction of the final 
uncertainty on $|V_{cb}|$ from 5.1$\times 10^{-4}$ to 4.8$\times 10^{-4}$. This is mostly due to the relevance of the theoretical
uncertainties. 
\begin{table}
    \centering
    \begin{tabular}{cccccccccc}
    \hline
    &&&&&&&\\[-4.5mm]
    &$m_b^{\rm kin}$ & $\overline{m}_c$ & $\mu_\pi^2$ & $\mu_G^2$ & $\rho_D^3$ & $\rho_{LS}^3$ & \!\!\!\!$10^2\text{BR}_{c\ell\nu}$ \!\! & \!\!\!\!$10^3 |V_{cb}|$ & \!\!\!\!
    \small{$\chi^2_{\rm min} (/\text{dof}$})\\
    &&&&&&&\\[-4.5mm]
    \hline \hline
     &&&&&&&\\[-4.5mm]
    without  & 4.573 & 1.092 & 0.477 & 0.306 & 0.185 & $-0.130$ & 10.66 & 42.16 & 22.3\\
    $q^2$-moments &0.012 & 0.008 & 0.056 & 0.050 & 0.031 & \phantom{$-$}0.092 & \phantom{1}0.15 & \phantom{4}0.51&0.474 \\
    \hline
    \hline
    &&&&&&&\\[-4.5mm]
    \multirow{ 2}{*}{Belle II} & 4.573 & 1.092 & 0.460 & 0.303 & 0.175 & $-0.118$ & 10.65 & 42.08 & 26.4\\
    &0.012 & 0.008 & 0.044 & 0.049 & 0.020 & \phantom{$-$}0.090 & \phantom{1}0.15 & \phantom{4}0.48& 0.425\\
    \hline
    \hline
    &&&&&&&\\[-4.5mm]
    \multirow{2}{*}{Belle} & 4.572 & 1.092 & 0.434 & 0.302 & 0.157 & $-0.100$ & 10.64 & 41.96 & 28.1\\
    &0.012 & 0.008 & 0.043 & 0.048 & 0.020 & \phantom{$-$}0.089 & \phantom{1}0.15 & \phantom{4}0.48& 0.476\\
    \hline
    \hline
     &&&&&&&\\[-4.5mm]
    Belle \& & 4.572 & 1.092 & 0.449 & 0.301 & 0.167 & $-0.109$ & 10.65 & 42.02 & 41.3\\
    Belle II &0.012 & 0.008 & 0.042 & 0.048 & 0.018 & \phantom{$-$}0.089 & \phantom{1}0.15 & \phantom{4}0.48& 0.559\\
    \hline
    \end{tabular}
    \caption{\small Global fit results with and without the $q^2$ moments from Belle/Belle II for $\mu_s = m_b/2$ and $\mu_c = 2~\text{GeV}$. All parameters are in GeV at the appropriate power and all, except $m_c$ , in the kinetic scheme at $\mu_k = 1$ GeV. The first row shows the central values and the second row the uncertainties. The first case corresponds to the default fit of \cite{Bordone:2021oof}.}
    \label{tab:fitBelleBelleII}
\end{table}
The analogue of Fig.~\ref{fig:ploterrors} with the parameters resulting from the fit including Belle and Belle II data is presented in Fig.~\ref{fig:ploterrorsq2}. We observe a clear reduction of the parametric uncertainty, mostly due to the improved determination of $\rho_D^3$.

\begin{figure}
    \centering
    \includegraphics[width=\textwidth]{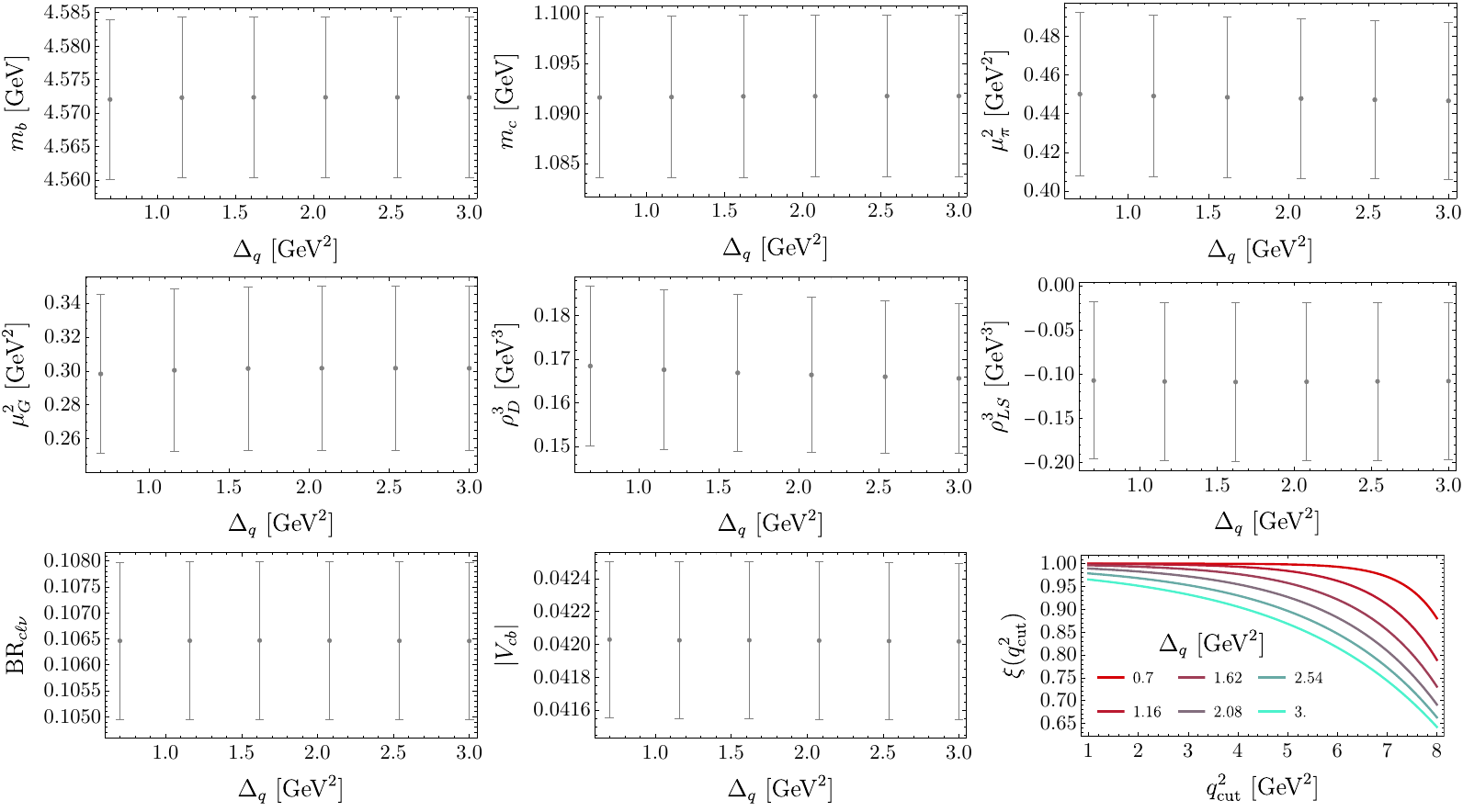}
    \caption{\small Results of global fits performed using different values of $\Delta_q$ in (\ref{eq:xi}). In the bottom-right panel we show the dependence of $\xi$ on $q^2_{\rm cut}$ for different values of $\Delta_q$.}
    \label{fig:Deltaq}
\end{figure}

We have performed a number of other fits, changing the scales  and selecting different subsets of data.
In particular, we study the dependence on the model of theoretical correlations by
varying  $\Delta_q$ in  between 0.7 and 3 GeV$^2$.   The results of the global fits including both Belle and Belle II data are shown in Fig.~\ref{fig:Deltaq}: they  depend very little on the choice for  $\Delta_q$.
As can be seen from (\ref{eq:xi})  the value of $\bar{q}^2$ controls the region in $q^2_{\rm cut}$
where the correlation between adjacent measurements starts to decrease because of fast growing higher order effects. Values of $\bar{q}^2$ lower than 9\,GeV$^2$ would lead to $\xi(q^2_{\rm cut})$ similar to those obtained with large $\Delta_q$, while values of $\bar{q}^2$ higher than 9\,GeV$^2$ appear unjustified.

\begin{figure}
    \centering
    \includegraphics[width=\textwidth]{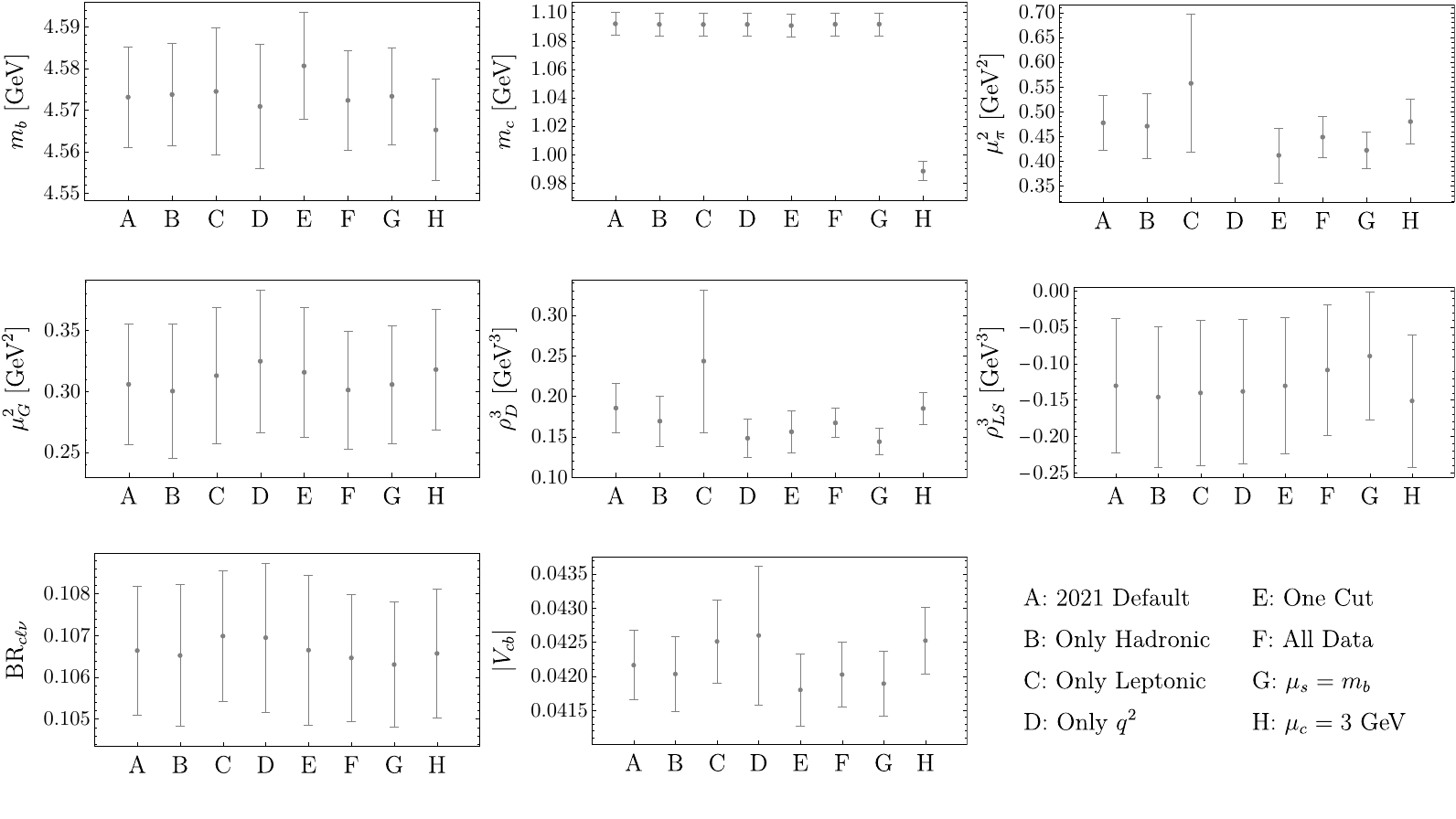}
    \caption{Fit results for different data sets (A-F), different choice of $\mu_s$ (G) and of the $\overline{\text{MS}}$ scale for the charm mass (H). The fit F corresponds to the last row of Table~\ref{tab:fitBelleBelleII}.}
    \label{fig:fits}
\end{figure}

The results of fits with various subsets of data are shown in Fig.~\ref{fig:fits}.
The fits with only hadronic moments and only $q^2$-moments also include the measurements of the branching fraction at different values of the cut on $E_\ell$ in order to determine $|V_{cb}|$. However, even including the branching fraction measurements, the only $q^2$-moments fit is unable to meaningfully constrain 
 $\mu_\pi^2$, as these moments are insensitive to this parameter. As a consequence, the result for $|V_{cb}|$ also suffers from a large uncertainty, as can be seen in Fig.~\ref{fig:fits}.
We also show the results of fits performed 
setting $\mu_s=m_b$ and $\mu_c=3$\,GeV. While there are differences, especially in the values of $\mu_\pi^2$ and
$\rho_D^3$, all fits are consistent within uncertainties. 
In order to test the importance of the inclusion of data measured with different values of the cuts on $E_\ell$ and $q^2$,
we also perform a fit using only a single cut for each leptonic, hadronic or $q^2$-moment. 
We select $q^2_{\rm cut}=5.5$ GeV$^2$, and the cut on $E_\ell$ closest to 1~GeV, depending on the experiment (we exclude Delphi's results, obtained without a cut on $E_\ell$), and use the Belle and BaBar measurements of the branching fraction at the lowest value $E_{\rm cut}=0.6~{\rm GeV}$.
This fit does not depend at all on the modelling of the theoretical correlations in the $q^2$-moments discussed above and only minimally on the modelling of the correlations in leptonic and hadronic moments. The results
are perfectly compatible with those in Table \ref{tab:fitBelleBelleII} and only slightly less precise: the final result is  $|V_{cb}|=41.80(53)\, 10^{-3}$. We conclude that, 
with the inclusion of $q^2$-moments, using 
moments at different cuts adds little to the global fit.

\subsection{Update of the semileptonic fit}
Thus far we have used the same inputs adopted in the default analysis of \cite{Bordone:2021oof}. 
In order to provide our final results we update the lattice QCD constraints on the $b$ and $c$ quark masses using the latest FLAG review \cite{FlavourLatticeAveragingGroupFLAG:2021npn}. The new FLAG $N_f=2+1+1$ heavy quark mass  averages are
\begin{equation}
\overline{m}_b^{(4)}(\overline{m}_b)=4.203(11)\, {\rm GeV}\,, \qquad \overline{m}_c^{(4)}(3\,{\rm GeV})=0.989(10)\,{\rm GeV}\,,
\label{eq:masses}
\end{equation}
where we have indicated the number of active quark flavours, which has to be taken into account 
in the conversion to the kinetic scheme. Converting $\overline{m}_b^{(4)}(\overline{m}_b)$ to 
$\overline{m}_b^{(5)}(\overline{m}_b) = 4.196(11)~\text{GeV}$
and then using the three loop results of \cite{Fael:2020iea,Fael:2020tow} (scheme B) we obtain 
the kinetic mass of the $b$ quark
\begin{equation}
m_b(1\;{\rm GeV}) = 4.562(18)~\text{GeV}\,.  
\end{equation}
Concerning the charm mass, we observe that the latest FLAG average has a larger uncertainty than in 2021, due to 
tensions between different determinations. Our default input is $\overline{m}_c(2\;\rm GeV)=1.094(11)~\text{GeV}$, obtained evolving $\overline{m}_c$ in (\ref{eq:masses}) from 
3 to 2 GeV. For $\as^{(5)}(M_Z) $ we  use the PDG value 0.1179(9) \cite{ParticleDataGroup:2022pth} and we keep the same constraints $\mu_G^2(m_b)=0.35(7)$GeV$^2$ and 
$\rho_{LS}^3=-0.15(10)$GeV$^3$ employed in \cite{Bordone:2021oof}.

The QED corrections to the leptonic moments have been recently computed in Ref.~\cite{Bigi:2023cbv}, where
small but non-negligible differences have been found with respect to the BaBar estimate based on PHOTOS.
We have investigated the importance of these differences in the context of the global fit. Let us illustrate our procedure with the example of the branching fraction  measured for $E_\ell>E_{\rm cut}$, $R(E_{\rm cut})$. 
 BaBar has measured \cite{BaBar:2004bij,BaBar:2009zpz} a photon inclusive branching fraction, $R_{\rm incl}(E_{\rm cut})$ and estimated the leading logarithmic soft-photon QED contribution $\Delta R(E_{\rm cut})$
 using PHOTOS \cite{Barberio:1993qi}. The 
 QED-subtracted branching ratio that we want to compare with our QCD-only theoretical predictions
 is therefore 
\begin{equation}
    R_{\rm QCD}(E_{\rm cut}) = R_{\rm incl}(E_{\rm cut})- \Delta R(E_{\rm cut})\,.
\end{equation}
The QED contribution $\Delta R(E_{\rm cut})$ have been computed in~\cite{Bigi:2023cbv} including the complete virtual contributions,
electroweak effects, and leading logarithmic power suppressed terms. In order to employ this new calculation
in place of the PHOTOS estimate we  write 
\begin{equation}
    R_{\rm QCD}^{\rm BaBar}+\Delta R^{\rm BaBar} = R_{\rm incl}\,, \qquad R_{\rm QCD}^{\rm new}+\Delta R^{\rm new} = R_{\rm incl}\,,
\end{equation}
where $\Delta R^{\rm new}$ is the result of the calculation of \cite{Bigi:2023cbv}
and $R_{\rm QCD}^{\rm new}$ is a more precise value for the QED-subtracted branching fraction. Hence we can express the correction in the form of a multiplicative factor $\zeta_{\rm QED}$ 
\begin{equation}
    R_{\rm QCD}^{\rm new} = \biggl[ 1+\frac{\Delta R^{\rm BaBar}-\Delta R^{\rm new}}{R_{\rm QCD}^{\rm BaBar}}\biggr]R_{\rm QCD}^{\rm BaBar} \equiv \zeta_{\rm QED} \, R_{\rm QCD}^{\rm BaBar}\,.
\end{equation}
From Ref.~\cite{Bigi:2023cbv} we obtain the values
\begin{equation}
    \zeta_{\rm QED}(0.6~\text{GeV}) = 0.9918\,, \qquad \zeta_{\rm QED}(1.2~\text{GeV}) = 0.9969\,, \qquad \zeta_{\rm QED}(1.5~\text{GeV}) = 1.0010\,,\nonumber
\end{equation}
which we use to correct BaBar results for the branching fraction. The first number is particularly important
because the branching fraction at the lowest $E_{\rm cut}$ drives the determination of the inclusive semileptonic branching fraction in the fits and because its sizeable $-0.8\%$ shift  may affect  the $|V_{cb}|$ determination in a visible way. We do not 
change the uncertainties and correlations given by BaBar.
We have proceeded in the same way with the leptonic moments measured by BaBar, and found that the changes in the fit are minimal. 
The Belle measurement of the leptonic moments \cite{Belle:2006kgy} subtracts the QED effects in a way  similar to what done by BaBar, but their paper does not provide the size of the subtraction. We are therefore unable to improve the QED treatment on the Belle data. Overall, the modified treatment of QED corrections leads to a $-0.23\%$ change in $|V_{cb}|$.

Finally, we briefly comment on the calculation of $O(\as^3)$ contributions to the semileptonic moments
at zero cuts performed in \cite{Fael:2022frj}. We have checked that the $O(\as^3)$ contributions
in the kinetic scheme given in \cite{Fael:2022frj} are generally well within our estimate of the theoretical uncertainty. The only exception appears to be the third hadronic central moment, where our $\sim 15\% $ uncertainty falls short of an  $O(\as^3)$ contribution exceeding 25\%. We therefore increase the theoretical uncertainty of the third hadronic moments for the values of $E_{\rm cut}$ where it is lower than $ 30\%$.
This affects mostly the third hadronic moment measured by Delphi \cite{DELPHI:2005mot}, which has an experimental uncertainty of about 20\% and favours a low $\rd$, and results in an increase of $\sim 0.008\;\rm GeV^3$ of the central value of $\rd$ in the fit.

\begin{figure}
    \centering
    \subfloat{\includegraphics[width=0.485\textwidth]{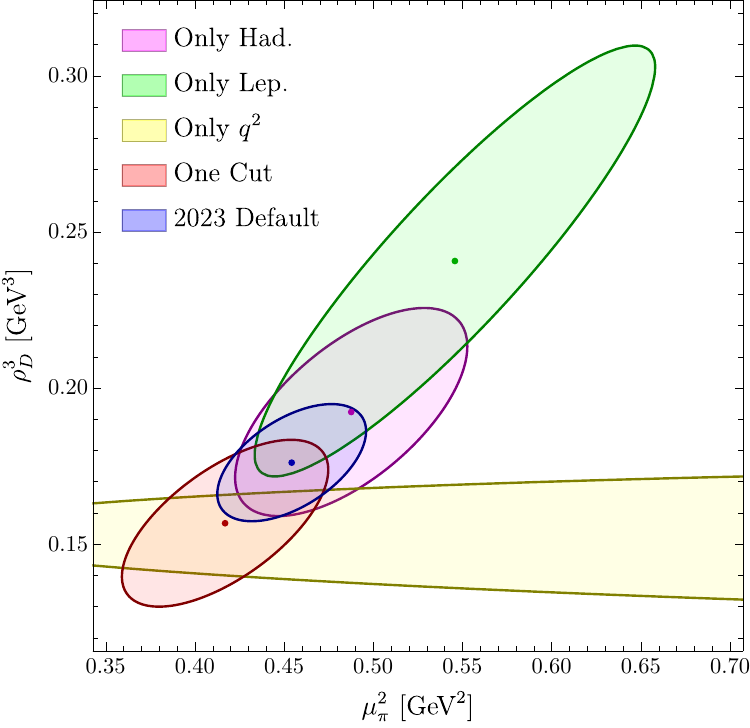}}$~$
    \subfloat{\includegraphics[width=0.495\textwidth]{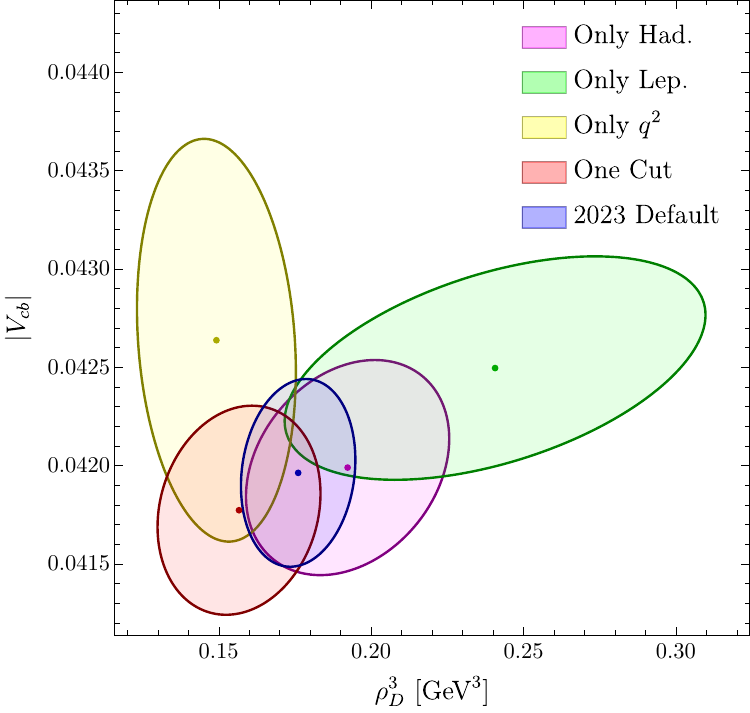}}
    \caption{\small Regions of $\Delta\chi^2 \leq 1$ in the 2D planes $(\mu_\pi^2,\rho_D^3)$ (left) and $(\rho_D^3,|V_{cb}|)$ (right). The dots stand for the points at $\Delta \chi^2=0$.}
    \label{fig:ellipses}
\end{figure}

Our final results are summarised in Table~\ref{tab:2}, where we present 
a global fit to hadronic, leptonic and $q^2$-moments that employs the updated heavy quark masses,
an enlarged theory uncertainty for the third hadronic moment, and includes, for the BaBar measurements,  the QED effects computed in \cite{Bigi:2023cbv}.
The changes with respect to the global fit (last row) of Table \ref{tab:fitBelleBelleII} are 
minor and mostly concern the determination of the branching fraction and  a $-0.1\%$ shift of $|V_{cb}|$.
In Fig.~\ref{fig:ellipses} we show the regions of $\Delta\chi^2<1$ in the 2D planes $(\mu_\pi^2,\rho_D^3)$ and $(\rho_D^3,|V_{cb}|)$, for the sets of data B-F of Fig.~\ref{fig:fits} after the various updates  discussed in this section.

\begin{table}[t]
  \begin{center} 
  \begin{tabular}{cccccccc}
  \hline
 $m_b^{\rm kin}$ & $\overline m_c(2\,\rm GeV)$   &  $\mupi $ &$\mug(m_b)$ & $\rd(m_b)$ & $\rls$  & ${\rm BR}_{c\ell\nu}$ & $10^3|V_{cb}|$ \\ 
 \hline
     4.573 & 1.090 & 0.454 & 0.288 & 0.176 & $-0.113$ & 10.63 & 41.97 \\  
    0.012 & 0.010 & 0.043 & 0.049 & 0.019 &  0.090 &  \;0.15 & \; 0.48\\ \hline
    1 & 0.380 & -0.219 & 0.557 & -0.013 & -0.172 & -0.063 & -0.428 \\
  & 1 & 0.005 & -0.235 & -0.051 & 0.083 & 0.030 & 0.071 \\
  &  & 1 & -0.083 & 0.537 & 0.241 & 0.140 & 0.335 \\
  &  &  & 1 & -0.247 & 0.010 & 0.007 & -0.253 \\
  &  &  &  & 1 & -0.023 & 0.023 & 0.140 \\
  &  &  &  &  & 1 & -0.011 & 0.060 \\
  &  &  &  &  &  & 1 & 0.696 \\
  &  &  &  &  &  &  & 1 \\
    \hline
   \end{tabular} 
   \end{center}
       \caption{\small Results of the updated fit in our default scenario ($\mu_c=2~\text{GeV}, \mu_s=m_b/2$).
All parameters are in GeV at the appropriate power and all, except $m_c$, in the kinetic scheme at $\mu_k=1$ GeV. The first and second rows give central values and uncertainties, the correlation matrix follows. $\chi^2_{\rm min} = 40.4$ and $\chi^2_{\rm min}/\text{dof}=0.546$.}
\label{tab:2} 
\end{table}

\section{Summary and outlook}
\label{sec:conc}
The recent measurements of the $q^2$-moments by Belle and Belle II \cite{Belle:2021idw, Belle-II:2022fug}
has opened new opportunities for the study of inclusive semileptonic $B$ decays. In this paper 
we have presented the results of a new calculation of the moments of the $q^2$ spectrum in inclusive semileptonic $B$ decays that includes contributions up to $O(\as^2\beta_0)$ and $O(\as\Lambda_{\rm QCD}^3/m_b^3)$.
In particular,  we have reproduced many of the results presented in Refs.~\cite{Fael:2018vsp,Mannel:2021zzr}
and computed for the first time the BLM corrections $O(\as^2\beta_0)$ to the $q^2$-moments.
If we employ the results of the default fit of \cite{Bordone:2021oof} as inputs, our predictions for the 
central moments of the $q^2$ spectrum are in excellent agreement with Belle II data \cite{Belle-II:2022fug},
while there is a mild tension with Belle data \cite{Belle:2021idw} in the case of the second and third central moments. As a matter of fact, the Belle and Belle II for those moments differ by about $2\sigma$. 

The inclusion of the $q^2$-moments in the global fit confirms the above  picture. The $q^2$-moments lower slightly the value of $\rd(m_b)$ by half a $\sigma$ and that  of $|V_{cb}|$ by a fraction of a $\sigma$,  decreasing the final uncertainty on them from $0.031$ to $0.018$\,GeV$^3$ and from 0.51$\times 10^{-3}$ to 0.48 $\times 10^{-3}$, respectively. Because of its 
correlation with $\rd$, the determination of $\mupi$ also benefit from the new data, with the uncertainty
going down from 0.056 to 0.042\,GeV$^2$.
We have also included the results of the new calculation of QED and electroweak effects on 
the lepton energy spectrum and moments \cite{Bigi:2023cbv}. Applying them to the BaBar data only, 
they lower the values of the branching fraction and of $|V_{cb}|$ by about 0.23\%. 
Our final result for $|V_{cb}|$, obtained updating the input charm and bottom masses and increasing 
the uncertainty on the hadronic moments, is
\be
|V_{cb}|= (41.97\pm 0.27_{exp} \pm 0.31_{th} \pm 0.25_\Gamma) \times 10^{-3}= (41.97\pm 0.48) \times 10^{-3}\, .
\ee
This is still in  tension with most estimates based on the Belle and BaBar measurements of exclusive decay $B\to D^* \ell\nu$ \cite{Gambino:2019sif,FermilabLattice:2021cdg,Aoki:2023qpa, Harrison:2023dzh,Bernlochner:2022ywh,Ray:2023xjn,Martinelli:2023fwm}, but agrees  well with the very recent Belle and Belle II results 
\cite{Belle-II:2023okj,Belle:2023bwv}
and with analyses of $B\to D \ell\nu$ \cite{Belle:2015pkj,Bigi:2016mdz}. Interestingly, we also find that a global fit to moments measured at a single cut on $E_\ell$ and $q^2$, which minimally depends on the correlations among theory errors, gives very similar results. This corroborates our study of the dependence on 
the modelling of theory correlations.

Further improvements of the inclusive determination of $|V_{cb}|$ may come from new and more precise 
measurements of the leptonic and hadronic moments  at Belle II, which could also measure the Forward-Backward 
asymmetry and related observables for the first time, bringing a new sensitivity to $\mu_G^2$ to the fits \cite{Turczyk:2016kjf, Herren:2022spb}.  The new measurements should be able to improve the treatment of QED corrections using the results of \cite{Bigi:2023cbv}. It will be useful to investigate the higher power contributions of
$O(\Lambda_{\rm QCD}^4/m_b^4, \Lambda_{\rm QCD}^5/(m_c^2 m_b^3), \Lambda_{\rm QCD}^5/m_b^5)$ in the $q^2$-moments, in analogy to what has been done in \cite{Gambino:2016jkc} for the hadronic and leptonic moments. As far as perturbative corrections are 
concerned, a complete $O(\as^2)$ calculation of the $q^2$-moments at arbitrary $q^2_{\rm cut}$ is feasible and necessary. The poor convergence of the perturbative series for the third hadronic moments
observed at $O(\as^3)$ in \cite{Fael:2022frj} should also be investigated.
In the longer term, we expect lattice calculation of the inclusive semileptonic $B$ decays \cite{Gambino:2020crt,Gambino:2022dvu, Barone:2023tbl}
to validate and complement the OPE calculations.

\acknowledgments{We thank Martin Jung for  discussions and suggestions. We also thank the authors of Ref.~\cite{Bernlochner:2022ucr} for providing us with their combination of Belle and Belle II $q^2$-moments results, 
Matteo Fael and Keri Vos for checking  Tables 1 and 2, and 
Martin Beneke and Florian~Bernlochner for useful discussions and explanations. The work of P.G. is supported in part by the Italian Ministry of Research (MIUR) under the grants PRIN 20172LNEEZ and 2022N4W8WR, by the Excellence Cluster ORIGINS which is funded by the Deutsche Forschungsgemeinschaft (DFG, German Research Foundation) under Germany's Excellence Strategy – EXC-2094 – 390783311, and by the Deutsche Forschungsgemeinschaft
(DFG, German Research Foundation) through the Sino-German Collaborative Research Center TRR110 "Symmetries and the Emergence of Structure in QCD" (DFG Project-ID 196253076, NSFC Grant No. 12070131001, - TRR 110).}

\bibliographystyle{JHEP}
\bibliography{refs}

\providecommand{\href}[2]{#2}\begingroup\raggedright\begin{thebibliography}{10}

\bibitem{BaBar:2004bij}
{\scshape BaBar} collaboration, B.~Aubert et~al., \emph{{Measurement of the
  electron energy spectrum and its moments in inclusive $B \to X e \nu$
  decays}}, \href{http://dx.doi.org/10.1103/PhysRevD.69.111104}{\emph{Phys.
  Rev. D} {\bf 69} (2004) 111104},
  [\href{http://arxiv.org/abs/hep-ex/0403030}{{\tt hep-ex/0403030}}].

\bibitem{CLEO:2004bqt}
{\scshape CLEO} collaboration, S.~E. Csorna et~al., \emph{{Moments of the B
  meson inclusive semileptonic decay rate using neutrino reconstruction}},
  \href{http://dx.doi.org/10.1103/PhysRevD.70.032002}{\emph{Phys. Rev. D} {\bf
  70} (2004) 032002}, [\href{http://arxiv.org/abs/hep-ex/0403052}{{\tt
  hep-ex/0403052}}].

\bibitem{CDF:2005xlh}
{\scshape CDF} collaboration, D.~Acosta et~al., \emph{{Measurement of the
  moments of the hadronic invariant mass distribution in semileptonic $B$
  decays}}, \href{http://dx.doi.org/10.1103/PhysRevD.71.051103}{\emph{Phys.
  Rev. D} {\bf 71} (2005) 051103},
  [\href{http://arxiv.org/abs/hep-ex/0502003}{{\tt hep-ex/0502003}}].

\bibitem{DELPHI:2005mot}
{\scshape DELPHI} collaboration, J.~Abdallah et~al., \emph{{Determination of
  heavy quark non-perturbative parameters from spectral moments in semileptonic
  B decays}}, \href{http://dx.doi.org/10.1140/epjc/s2005-02406-7}{\emph{Eur.
  Phys. J. C} {\bf 45} (2006) 35--59},
  [\href{http://arxiv.org/abs/hep-ex/0510024}{{\tt hep-ex/0510024}}].

\bibitem{Belle:2006jtu}
{\scshape Belle} collaboration, C.~Schwanda et~al., \emph{{Moments of the
  Hadronic Invariant Mass Spectrum in $B \to X_c \ell \nu$ Decays at {BELLE}}},
  \href{http://dx.doi.org/10.1103/PhysRevD.75.032005}{\emph{Phys. Rev. D} {\bf
  75} (2007) 032005}, [\href{http://arxiv.org/abs/hep-ex/0611044}{{\tt
  hep-ex/0611044}}].

\bibitem{Belle:2006kgy}
{\scshape Belle} collaboration, P.~Urquijo et~al., \emph{{Moments of the
  electron energy spectrum and partial branching fraction of B
  ---\ensuremath{>} X(c) e nu decays at Belle}},
  \href{http://dx.doi.org/10.1103/PhysRevD.75.032001}{\emph{Phys. Rev. D} {\bf
  75} (2007) 032001}, [\href{http://arxiv.org/abs/hep-ex/0610012}{{\tt
  hep-ex/0610012}}].

\bibitem{BaBar:2009zpz}
{\scshape BaBar} collaboration, B.~Aubert et~al., \emph{{Measurement and
  interpretation of moments in inclusive semileptonic decays anti-B
  ---\ensuremath{>} X(c) l- anti-nu}},
  \href{http://dx.doi.org/10.1103/PhysRevD.81.032003}{\emph{Phys. Rev. D} {\bf
  81} (2010) 032003}, [\href{http://arxiv.org/abs/0908.0415}{{\tt 0908.0415}}].

\bibitem{Bauer:2004ve}
C.~W. Bauer, Z.~Ligeti, M.~Luke, A.~V. Manohar and M.~Trott, \emph{{Global
  analysis of inclusive B decays}},
  \href{http://dx.doi.org/10.1103/PhysRevD.70.094017}{\emph{Phys. Rev. D} {\bf
  70} (2004) 094017}, [\href{http://arxiv.org/abs/hep-ph/0408002}{{\tt
  hep-ph/0408002}}].

\bibitem{Buchmuller:2005zv}
O.~Buchmuller and H.~Flacher, \emph{{Fit to moment from B ---\ensuremath{>}
  X(c) l anti-nu and B ---\ensuremath{>} X(s) gamma decays using heavy quark
  expansions in the kinetic scheme}},
  \href{http://dx.doi.org/10.1103/PhysRevD.73.073008}{\emph{Phys. Rev. D} {\bf
  73} (2006) 073008}, [\href{http://arxiv.org/abs/hep-ph/0507253}{{\tt
  hep-ph/0507253}}].

\bibitem{Gambino:2013rza}
P.~Gambino and C.~Schwanda, \emph{{Inclusive semileptonic fits, heavy quark
  masses, and $V_{cb}$}},
  \href{http://dx.doi.org/10.1103/PhysRevD.89.014022}{\emph{Phys. Rev. D} {\bf
  89} (2014) 014022}, [\href{http://arxiv.org/abs/1307.4551}{{\tt 1307.4551}}].

\bibitem{Alberti:2014yda}
A.~Alberti, P.~Gambino, K.~J. Healey and S.~Nandi, \emph{{Precision
  Determination of the Cabibbo-Kobayashi-Maskawa Element $V_{cb}$}},
  \href{http://dx.doi.org/10.1103/PhysRevLett.114.061802}{\emph{Phys. Rev.
  Lett.} {\bf 114} (2015) 061802}, [\href{http://arxiv.org/abs/1411.6560}{{\tt
  1411.6560}}].

\bibitem{Bordone:2021oof}
M.~Bordone, B.~Capdevila and P.~Gambino, \emph{{Three loop calculations and
  inclusive Vcb}},
  \href{http://dx.doi.org/10.1016/j.physletb.2021.136679}{\emph{Phys. Lett. B}
  {\bf 822} (2021) 136679}, [\href{http://arxiv.org/abs/2107.00604}{{\tt
  2107.00604}}].

\bibitem{HeavyFlavorAveragingGroup:2022wzx}
{\scshape Heavy Flavor Averaging Group, HFLAV} collaboration, Y.~S. Amhis
  et~al., \emph{{Averages of b-hadron, c-hadron, and \ensuremath{\tau}-lepton
  properties as of 2021}},
  \href{http://dx.doi.org/10.1103/PhysRevD.107.052008}{\emph{Phys. Rev. D} {\bf
  107} (2023) 052008}, [\href{http://arxiv.org/abs/2206.07501}{{\tt
  2206.07501}}].

\bibitem{Fael:2020tow}
M.~Fael, K.~Sch\"onwald and M.~Steinhauser, \emph{{Third order corrections to
  the semileptonic b\textrightarrow{}c and the muon decays}},
  \href{http://dx.doi.org/10.1103/PhysRevD.104.016003}{\emph{Phys. Rev. D} {\bf
  104} (2021) 016003}, [\href{http://arxiv.org/abs/2011.13654}{{\tt
  2011.13654}}].

\bibitem{Fael:2018vsp}
M.~Fael, T.~Mannel and K.~Keri~Vos, \emph{{$V_{cb}$ determination from
  inclusive $b \to c$ decays: an alternative method}},
  \href{http://dx.doi.org/10.1007/JHEP02(2019)177}{\emph{JHEP} {\bf 02} (2019)
  177}, [\href{http://arxiv.org/abs/1812.07472}{{\tt 1812.07472}}].

\bibitem{Luke:1992cs}
M.~E. Luke and A.~V. Manohar, \emph{{Reparametrization invariance constraints
  on heavy particle effective field theories}},
  \href{http://dx.doi.org/10.1016/0370-2693(92)91786-9}{\emph{Phys. Lett. B}
  {\bf 286} (1992) 348--354}, [\href{http://arxiv.org/abs/hep-ph/9205228}{{\tt
  hep-ph/9205228}}].

\bibitem{Mannel:2010wj}
T.~Mannel, S.~Turczyk and N.~Uraltsev, \emph{{Higher Order Power Corrections in
  Inclusive B Decays}},
  \href{http://dx.doi.org/10.1007/JHEP11(2010)109}{\emph{JHEP} {\bf 11} (2010)
  109}, [\href{http://arxiv.org/abs/1009.4622}{{\tt 1009.4622}}].

\bibitem{Belle:2021idw}
{\scshape Belle} collaboration, R.~van Tonder et~al., \emph{{Measurements of
  $q^2$ Moments of Inclusive $B \rightarrow X_c \ell^+ \nu_{\ell}$ Decays with
  Hadronic Tagging}},
  \href{http://dx.doi.org/10.1103/PhysRevD.104.112011}{\emph{Phys. Rev. D} {\bf
  104} (2021) 112011}, [\href{http://arxiv.org/abs/2109.01685}{{\tt
  2109.01685}}].

\bibitem{Belle-II:2022fug}
{\scshape Belle-II} collaboration, \emph{{Measurement of Lepton Mass Squared
  Moments in $B \to X_c \ell \bar \nu_{\ell}$ Decays with the Belle II
  Experiment}},  \href{http://arxiv.org/abs/2205.06372}{{\tt 2205.06372}}.

\bibitem{Bernlochner:2022ucr}
F.~Bernlochner, M.~Fael, K.~Olschewsky, E.~Persson, R.~van Tonder, K.~K. Vos
  et~al., \emph{{First extraction of inclusive V$_{cb}$ from q$^{2}$ moments}},
  \href{http://dx.doi.org/10.1007/JHEP10(2022)068}{\emph{JHEP} {\bf 10} (2022)
  068}, [\href{http://arxiv.org/abs/2205.10274}{{\tt 2205.10274}}].

\bibitem{Gambino:2004qm}
P.~Gambino and N.~Uraltsev, \emph{{Moments of semileptonic B decay
  distributions in the 1/m(b) expansion}},
  \href{http://dx.doi.org/10.1140/epjc/s2004-01671-2}{\emph{Eur. Phys. J. C}
  {\bf 34} (2004) 181--189}, [\href{http://arxiv.org/abs/hep-ph/0401063}{{\tt
  hep-ph/0401063}}].

\bibitem{Blok:1993va}
B.~Blok, L.~Koyrakh, M.~A. Shifman and A.~I. Vainshtein, \emph{{Differential
  distributions in semileptonic decays of the heavy flavors in QCD}},
  \href{http://dx.doi.org/10.1103/PhysRevD.50.3572}{\emph{Phys. Rev. D} {\bf
  49} (1994) 3356}, [\href{http://arxiv.org/abs/hep-ph/9307247}{{\tt
  hep-ph/9307247}}].

\bibitem{Manohar:1993qn}
A.~V. Manohar and M.~B. Wise, \emph{{Inclusive semileptonic B and polarized
  Lambda(b) decays from QCD}},
  \href{http://dx.doi.org/10.1103/PhysRevD.49.1310}{\emph{Phys. Rev. D} {\bf
  49} (1994) 1310--1329}, [\href{http://arxiv.org/abs/hep-ph/9308246}{{\tt
  hep-ph/9308246}}].

\bibitem{Gremm:1996df}
M.~Gremm and A.~Kapustin, \emph{{Order 1/m(b)**3 corrections to B
  --\ensuremath{>} X(c) lepton anti-neutrino decay and their implication for
  the measurement of Lambda-bar and lambda(1)}},
  \href{http://dx.doi.org/10.1103/PhysRevD.55.6924}{\emph{Phys. Rev. D} {\bf
  55} (1997) 6924--6932}, [\href{http://arxiv.org/abs/hep-ph/9603448}{{\tt
  hep-ph/9603448}}].

\bibitem{Gambino:2016jkc}
P.~Gambino, K.~J. Healey and S.~Turczyk, \emph{{Taming the higher power
  corrections in semileptonic B decays}},
  \href{http://dx.doi.org/10.1016/j.physletb.2016.10.023}{\emph{Phys. Lett. B}
  {\bf 763} (2016) 60--65}, [\href{http://arxiv.org/abs/1606.06174}{{\tt
  1606.06174}}].

\bibitem{Aquila:2005hq}
V.~Aquila, P.~Gambino, G.~Ridolfi and N.~Uraltsev, \emph{{Perturbative
  corrections to semileptonic b decay distributions}},
  \href{http://dx.doi.org/10.1016/j.nuclphysb.2005.04.031}{\emph{Nucl. Phys. B}
  {\bf 719} (2005) 77--102}, [\href{http://arxiv.org/abs/hep-ph/0503083}{{\tt
  hep-ph/0503083}}].

\bibitem{Trott:2004xc}
M.~Trott, \emph{{Improving extractions of |V(cb)| and m(b) from the hadronic
  invariant mass moments of semileptonic inclusive B decay}},
  \href{http://dx.doi.org/10.1103/PhysRevD.70.073003}{\emph{Phys. Rev. D} {\bf
  70} (2004) 073003}, [\href{http://arxiv.org/abs/hep-ph/0402120}{{\tt
  hep-ph/0402120}}].

\bibitem{Alberti:2012dn}
A.~Alberti, T.~Ewerth, P.~Gambino and S.~Nandi, \emph{{Kinetic operator effects
  in $\bar{B}\to X_c l \nu$ at O($\alpha_s$)}},
  \href{http://dx.doi.org/10.1016/j.nuclphysb.2013.01.005}{\emph{Nucl. Phys. B}
  {\bf 870} (2013) 16--29}, [\href{http://arxiv.org/abs/1212.5082}{{\tt
  1212.5082}}].

\bibitem{Alberti:2013kxa}
A.~Alberti, P.~Gambino and S.~Nandi, \emph{{Perturbative corrections to power
  suppressed effects in semileptonic B decays}},
  \href{http://dx.doi.org/10.1007/JHEP01(2014)147}{\emph{JHEP} {\bf 01} (2014)
  147}, [\href{http://arxiv.org/abs/1311.7381}{{\tt 1311.7381}}].

\bibitem{Mannel:2021zzr}
T.~Mannel, D.~Moreno and A.~A. Pivovarov, \emph{{NLO QCD corrections to
  inclusive $b \rightarrow c \ell \bar{\nu}$decay spectra up to~$1/m_Q^3$}},
  \href{http://dx.doi.org/10.1103/PhysRevD.105.054033}{\emph{Phys. Rev. D} {\bf
  105} (2022) 054033}, [\href{http://arxiv.org/abs/2112.03875}{{\tt
  2112.03875}}].

\bibitem{Manohar:2010sf}
A.~V. Manohar, \emph{{Reparametrization Invariance Constraints on Inclusive
  Decay Spectra and Masses}},
  \href{http://dx.doi.org/10.1103/PhysRevD.82.014009}{\emph{Phys. Rev. D} {\bf
  82} (2010) 014009}, [\href{http://arxiv.org/abs/1005.1952}{{\tt 1005.1952}}].

\bibitem{Colangelo:2020vhu}
P.~Colangelo, F.~De~Fazio and F.~Loparco, \emph{{Inclusive semileptonic
  $\Lambda_{b}$ decays in the Standard Model and beyond}},
  \href{http://dx.doi.org/10.1007/JHEP11(2020)032}{\emph{JHEP} {\bf 11} (2020)
  032}, [\href{http://arxiv.org/abs/2006.13759}{{\tt 2006.13759}}].

\bibitem{Bigi:1996si}
I.~I.~Y. Bigi, M.~A. Shifman, N.~Uraltsev and A.~I. Vainshtein, \emph{{High
  power n of m(b) in beauty widths and n=5 ---\ensuremath{>} infinity limit}},
  \href{http://dx.doi.org/10.1103/PhysRevD.56.4017}{\emph{Phys. Rev. D} {\bf
  56} (1997) 4017--4030}, [\href{http://arxiv.org/abs/hep-ph/9704245}{{\tt
  hep-ph/9704245}}].

\bibitem{Czarnecki:1997sz}
A.~Czarnecki, K.~Melnikov and N.~Uraltsev, \emph{{NonAbelian dipole radiation
  and the heavy quark expansion}},
  \href{http://dx.doi.org/10.1103/PhysRevLett.80.3189}{\emph{Phys. Rev. Lett.}
  {\bf 80} (1998) 3189--3192}, [\href{http://arxiv.org/abs/hep-ph/9708372}{{\tt
  hep-ph/9708372}}].

\bibitem{Fael:2020iea}
M.~Fael, K.~Sch\"onwald and M.~Steinhauser, \emph{{Kinetic Heavy Quark Mass to
  Three Loops}},
  \href{http://dx.doi.org/10.1103/PhysRevLett.125.052003}{\emph{Phys. Rev.
  Lett.} {\bf 125} (2020) 052003}, [\href{http://arxiv.org/abs/2005.06487}{{\tt
  2005.06487}}].

\bibitem{Melnikov:2000qh}
K.~Melnikov and T.~v. Ritbergen, \emph{{The Three loop relation between the
  MS-bar and the pole quark masses}},
  \href{http://dx.doi.org/10.1016/S0370-2693(00)00507-4}{\emph{Phys. Lett. B}
  {\bf 482} (2000) 99--108}, [\href{http://arxiv.org/abs/hep-ph/9912391}{{\tt
  hep-ph/9912391}}].

\bibitem{Fael:2022frj}
M.~Fael, K.~Sch\"onwald and M.~Steinhauser, \emph{{A first glance to the
  kinematic moments of B \textrightarrow{}
  X$_{c}$\ensuremath{\ell}\ensuremath{\nu} at third order}},
  \href{http://dx.doi.org/10.1007/JHEP08(2022)039}{\emph{JHEP} {\bf 08} (2022)
  039}, [\href{http://arxiv.org/abs/2205.03410}{{\tt 2205.03410}}].

\bibitem{Gambino:2011cq}
P.~Gambino, \emph{{B semileptonic moments at NNLO}},
  \href{http://dx.doi.org/10.1007/JHEP09(2011)055}{\emph{JHEP} {\bf 09} (2011)
  055}, [\href{http://arxiv.org/abs/1107.3100}{{\tt 1107.3100}}].

\bibitem{FlavourLatticeAveragingGroupFLAG:2021npn}
{\scshape Flavour Lattice Averaging Group (FLAG)} collaboration, Y.~Aoki
  et~al., \emph{{FLAG Review 2021}},
  \href{http://dx.doi.org/10.1140/epjc/s10052-022-10536-1}{\emph{Eur. Phys. J.
  C} {\bf 82} (2022) 869}, [\href{http://arxiv.org/abs/2111.09849}{{\tt
  2111.09849}}].

\bibitem{ParticleDataGroup:2022pth}
{\scshape Particle Data Group} collaboration, R.~L. Workman et~al.,
  \emph{{Review of Particle Physics}},
  \href{http://dx.doi.org/10.1093/ptep/ptac097}{\emph{PTEP} {\bf 2022} (2022)
  083C01}.

\bibitem{Bigi:2023cbv}
D.~Bigi, M.~Bordone, P.~Gambino, U.~Haisch and A.~Piccione, \emph{{QED effects
  in inclusive semi-leptonic $B$ decays}},
  \href{http://arxiv.org/abs/2309.02849}{{\tt 2309.02849}}.

\bibitem{Barberio:1993qi}
E.~Barberio and Z.~Was, \emph{{PHOTOS: A Universal Monte Carlo for QED
  radiative corrections. Version 2.0}},
  \href{http://dx.doi.org/10.1016/0010-4655(94)90074-4}{\emph{Comput. Phys.
  Commun.} {\bf 79} (1994) 291--308}.

\bibitem{Gambino:2019sif}
P.~Gambino, M.~Jung and S.~Schacht, \emph{{The $V_{cb}$ puzzle: An update}},
  \href{http://dx.doi.org/10.1016/j.physletb.2019.06.039}{\emph{Phys. Lett. B}
  {\bf 795} (2019) 386--390}, [\href{http://arxiv.org/abs/1905.08209}{{\tt
  1905.08209}}].

\bibitem{FermilabLattice:2021cdg}
{\scshape Fermilab Lattice, MILC, Fermilab Lattice, MILC} collaboration,
  A.~Bazavov et~al., \emph{{Semileptonic form factors for $B\rightarrow D^*\ell
  \nu $ at nonzero recoil from $2+1$-flavor lattice QCD: Fermilab
  Lattice~and~MILC~Collaborations}},
  \href{http://dx.doi.org/10.1140/epjc/s10052-022-10984-9}{\emph{Eur. Phys. J.
  C} {\bf 82} (2022) 1141}, [\href{http://arxiv.org/abs/2105.14019}{{\tt
  2105.14019}}].

\bibitem{Aoki:2023qpa}
{\scshape JLQCD} collaboration, Y.~Aoki, B.~Colquhoun, H.~Fukaya, S.~Hashimoto,
  T.~Kaneko, R.~Kellermann et~al., \emph{{$B \to D^*\ell\nu_\ell$ semileptonic
  form factors from lattice QCD with M\"obius domain-wall quarks}},
  \href{http://arxiv.org/abs/2306.05657}{{\tt 2306.05657}}.

\bibitem{Harrison:2023dzh}
J.~Harrison and C.~T.~H. Davies, \emph{{$B \rightarrow D^*$ vector,
  axial-vector and tensor form factors for the full $q^2$ range from lattice
  QCD}},  \href{http://arxiv.org/abs/2304.03137}{{\tt 2304.03137}}.

\bibitem{Bernlochner:2022ywh}
F.~U. Bernlochner, Z.~Ligeti, M.~Papucci, M.~T. Prim, D.~J. Robinson and
  C.~Xiong, \emph{{Constrained second-order power corrections in HQET: R(D(*)),
  |Vcb|, and new physics}},
  \href{http://dx.doi.org/10.1103/PhysRevD.106.096015}{\emph{Phys. Rev. D} {\bf
  106} (2022) 096015}, [\href{http://arxiv.org/abs/2206.11281}{{\tt
  2206.11281}}].

\bibitem{Ray:2023xjn}
I.~Ray and S.~Nandi, \emph{{Test of new physics effects in $\bar{B} \to
  (D^{(*)}, \pi) \ell^-\bar{\nu}_{\ell}$ decays with heavy and light leptons}},
   \href{http://arxiv.org/abs/2305.11855}{{\tt 2305.11855}}.

\bibitem{Martinelli:2023fwm}
G.~Martinelli, S.~Simula and L.~Vittorio, \emph{{Updates on the determination
  of $\vert V_{cb} \vert$, $R(D^{*})$ and $\vert V_{ub} \vert/\vert V_{cb}
  \vert$}},  \href{http://arxiv.org/abs/2310.03680}{{\tt 2310.03680}}.

\bibitem{Belle-II:2023okj}
{\scshape Belle-II} collaboration, I.~Adachi et~al., \emph{{Determination of
  $|V_{cb}|$ using $\overline{B}^0\to D^{*+}\ell^-\bar\nu_\ell$ decays with
  Belle II}},  \href{http://arxiv.org/abs/2310.01170}{{\tt 2310.01170}}.

\bibitem{Belle:2023bwv}
{\scshape Belle} collaboration, M.~T. Prim et~al., \emph{{Measurement of
  differential distributions of
  B\textrightarrow{}D*\ensuremath{\ell}\ensuremath{\nu}\textasciimacron{}\ensuremath{\ell}
  and implications on |Vcb|}},
  \href{http://dx.doi.org/10.1103/PhysRevD.108.012002}{\emph{Phys. Rev. D} {\bf
  108} (2023) 012002}, [\href{http://arxiv.org/abs/2301.07529}{{\tt
  2301.07529}}].

\bibitem{Belle:2015pkj}
{\scshape Belle} collaboration, R.~Glattauer et~al., \emph{{Measurement of the
  decay $B\to D\ell\nu_\ell$ in fully reconstructed events and determination of
  the Cabibbo-Kobayashi-Maskawa matrix element $|V_{cb}|$}},
  \href{http://dx.doi.org/10.1103/PhysRevD.93.032006}{\emph{Phys. Rev. D} {\bf
  93} (2016) 032006}, [\href{http://arxiv.org/abs/1510.03657}{{\tt
  1510.03657}}].

\bibitem{Bigi:2016mdz}
D.~Bigi and P.~Gambino, \emph{{Revisiting $B\to D \ell \nu$}},
  \href{http://dx.doi.org/10.1103/PhysRevD.94.094008}{\emph{Phys. Rev. D} {\bf
  94} (2016) 094008}, [\href{http://arxiv.org/abs/1606.08030}{{\tt
  1606.08030}}].

\bibitem{Turczyk:2016kjf}
S.~Turczyk, \emph{{Additional Information on Heavy Quark Parameters from
  Charged Lepton Forward-Backward Asymmetry}},
  \href{http://dx.doi.org/10.1007/JHEP04(2016)131}{\emph{JHEP} {\bf 04} (2016)
  131}, [\href{http://arxiv.org/abs/1602.02678}{{\tt 1602.02678}}].

\bibitem{Herren:2022spb}
F.~Herren, \emph{{The forward-backward asymmetry and differences of partial
  moments in inclusive semileptonic $B$ decays}},
  \href{http://dx.doi.org/10.21468/SciPostPhys.14.2.020}{\emph{SciPost Phys.}
  {\bf 14} (2023) 020}, [\href{http://arxiv.org/abs/2205.03427}{{\tt
  2205.03427}}].

\bibitem{Gambino:2020crt}
P.~Gambino and S.~Hashimoto, \emph{{Inclusive Semileptonic Decays from Lattice
  QCD}}, \href{http://dx.doi.org/10.1103/PhysRevLett.125.032001}{\emph{Phys.
  Rev. Lett.} {\bf 125} (2020) 032001},
  [\href{http://arxiv.org/abs/2005.13730}{{\tt 2005.13730}}].

\bibitem{Gambino:2022dvu}
P.~Gambino, S.~Hashimoto, S.~M\"achler, M.~Panero, F.~Sanfilippo, S.~Simula
  et~al., \emph{{Lattice QCD study of inclusive semileptonic decays of heavy
  mesons}}, \href{http://dx.doi.org/10.1007/JHEP07(2022)083}{\emph{JHEP} {\bf
  07} (2022) 083}, [\href{http://arxiv.org/abs/2203.11762}{{\tt 2203.11762}}].

\bibitem{Barone:2023tbl}
A.~Barone, S.~Hashimoto, A.~J\"uttner, T.~Kaneko and R.~Kellermann,
  \emph{{Approaches to inclusive semileptonic B$_{(s)}$-meson decays from
  Lattice QCD}}, \href{http://dx.doi.org/10.1007/JHEP07(2023)145}{\emph{JHEP}
  {\bf 07} (2023) 145}, [\href{http://arxiv.org/abs/2305.14092}{{\tt
  2305.14092}}].

\end{thebibliography}\endgroup
\end{document}